\documentclass[article]{jss}

\usepackage{thumbpdf,lmodern}

\usepackage{framed}

\usepackage{comment}

\usepackage{amsfonts}
\usepackage{amsmath}
\usepackage{algorithm}
\usepackage{algpseudocode}
\usepackage{pbox}
\DeclareMathAlphabet\mathbfcal{OMS}{cmsy}{b}{n}
\usepackage{"./macro/GrandMacros"}
\usepackage{"./macro/Macro_BIO235"}
\def\real{{\mathbb R}}
\def\bone{{\bf 1}}
\newcommand{\la}{\langle}
\newcommand{\ra}{\rangle}
\author{Wenying Deng\\Harvard University
   \And Jeremiah Zhe Liu\\Harvard University
   \And Erin Lake\\Harvard University      
   \And Brent A. Coull\\Harvard University
   }
   \Plainauthor{Wenying Deng, Jeremiah Zhe Liu, Erin Lake, Brent A. Coull}

\title{\pkg{CVEK}: Robust Estimation and Testing for Nonlinear Effects using Kernel Machine Ensemble}
\Shorttitle{Cross-Validated Kernel Ensemble}

\Abstract{
The \proglang{R} package \pkg{CVEK} introduces a suite of flexible machine learning models and robust hypothesis tests for learning the joint nonlinear effects of multiple covariates in limited samples. It implements the \textit{Cross-validated Ensemble of Kernels} (CVEK)\citep{liu_robust_2017}, an ensemble-based kernel  machine learning method that adaptively learns the joint nonlinear effect of multiple covariates from data, and provides powerful hypothesis tests for both main effects of features and interactions among features. The R Package \pkg{CVEK} provides a flexible, easy-to-use implementation of CVEK, and offers a wide range of choices for the kernel family (for instance, polynomial, radial basis functions, Mat\'{e}rn, neural network, and others), model selection criteria,  ensembling method (averaging, exponential weighting, cross-validated stacking), and the type of hypothesis test (asymptotic or parametric bootstrap). Through extensive simulations we demonstrate the validity and robustness of this approach, and provide practical guidelines on how to design an estimation strategy for optimal performance in different data scenarios.
}

\Plainkeywords{robust hypothesis test, nonlinear effect, Gaussian process, CVEK, R, kernel ensemble}

\Address{
  Wenying Deng, Jeremiah Zhe Liu, Erin Lake, Brent A. Coull\\
  Department of Biostatistics\\
  Harvard University\\
  Cambridge, MA 02138 USA\\
  E-mail: \email{\{wdeng@g, zhl112@mail, eklake@hsph, bcoull@hsph\}.harvard.edu}
}
\begin{document}

%% -- Introduction -------------------------------------------------------------

%% - In principle "as usual".
%% - But should typically have some discussion of both _software_ and _methods_.
%% - Use \proglang{}, \pkg{}, and \code{} markup throughout the manuscript.
%% - If such markup is in (sub)section titles, a plain text version has to be
%%   added as well.
%% - All software mentioned should be properly \cite-d.
%% - All abbreviations should be introduced.
%% - Unless the expansions of abbreviations are proper names (like "Journal
%%   of Statistical Software" above) they should be in sentence case (like
%%   "generalized linear models" below).

\raggedright
\section[Introduction]{Introduction} \label{sec:intro}

In recent years, kernel machine methods have seen widespread application in biomedical studies for learning the complex, nonlinear effects of multivariate genetic or environmental exposures. Given data and feature $\{y, \bx\}$, practitioners are often interested in constructing a function $\hat{h}(\bx)$ that best describes the data generation mechanism $y = \mu + h(\bx)+\epsilon$. Further, given groups of features $\{\bx_1, \bx_2\} \subset \bx$, interest may focus on conducting a hypothesis test for either the overall effect of $\bx_1$ on $y$, or on the interaction effect between two feature groups $\bx_1$ and  $\bx_2$ based on $h$. 

Traditionally, kernel machine regression (KMR) handles this task by specifying a  kernel function $k(\bx, \bx')$ that gives rise to a large function space $\Hsc_k$, and estimates the data generation function $h$ through regularized regression. Using the estimated $\hat{h}$, hypothesis tests for feature effects can become simply variance component tests by interpreting the KMR as a linear mixed model \citep{liu_semiparametric_2007}. Given unlimited data and proper choice of kernel family, kernel machine regression enjoys a theoretical guarantee of learning arbitrary continuous target functions defined over a compact input space \citep{micchelli_universal_2006}, thereby inducing a valid hypothesis test with correct Type I error. 

In practice, however, the performance of KMR in limited samples is known to be extremely sensitive to the choices of the kernel function. To guarantee reasonable performance, $k$ must be selected carefully so that its mathematical properties reflect those of the data generating mechanism. Selecting an overly smooth kernel will result in an $\hat{h}$ that underfits the data, inducing an invalid hypothesis test with inflated Type I error. Selecting an overly flexible kernel function will lead to $\hat{h}$ that overfits the data, leading to an underpowered test. For most applications in epidemiology and the natural sciences, it is often difficult to specify a kernel family \textit{a priori} for a complex, nonlinear data generation mechanism $h$. This leads to challenges in both estimation and testing.

\pkg{CVEK} is an R package that provides a suite of robust estimation and testing procedures that adaptively learn the proper kernel function from the data through the use of kernel ensembling, thereby achieving unbiased effect estimation and valid hypothesis testing in limited samples. Specifically, \pkg{CVEK} implements the \textit{Cross-validated Ensemble of Kernels} (CVEK) \citep{liu_robust_2017, liu_cross-validated_2019}, an ensemble-based kernel machine learning procedure that automatically discerns the most appropriate kernel for the data using a cross-validated approach. \pkg{CVEK} offers a range of choices in base kernel families, model selection criteria, and ensemble methods, so the practitioner can flexibly design a modeling strategy for the data at hand. Section \ref{sec:cvek} introduces CVEK and such choices in detail. Section \ref{sec:test} presents \pkg{CVEK}'s implementation of asymptotic and bootstrap-based hypothesis tests for the overall effect of a single feature/group, as well as the interaction effect between features/groups. Section \ref{sec:package} provides a hands-on tutorial in learning and testing gene-environment interactions with \pkg{CVEK}. Section \ref{sec:simu} discusses the impact of choice of estimation strategy (i.e., choices of model selection criteria and ensemble strategy) on the performance of the resulting hypothesis test. This is demonstrated through a comprehensive simulation study that evaluates validity (i.e., Type I error) and power of the implemented tests using diverse modeling strategies under a wide range of data generation mechanisms.

%% -- Manuscript ---------------------------------------------------------------

%% - In principle "as usual" again.
%% - When using equations (e.g., {equation}, {eqnarray}, {align}, etc.
%%   avoid empty lines before and after the equation (which would signal a new
%%   paragraph.
%% - When describing longer chunks of code that are _not_ meant for execution
%%   (e.g., a function synopsis or list of arguments), the environment {Code}
%%   is recommended. Alternatively, a plain {verbatim} can a^\toplso be used.
%%   (For executed code see the next section.)

\raggedright
\section{Robust estimation using kernel ensemble via CVEK} 
\label{sec:cvek}

\subsection{Gaussian process regression}
Assume we observe data from $n$ independent subjects. For the $i^{th}$ subject, let $y_i$ be a continuous response, $\bx_i$ be the set of $p$ continuous features that have a potentially nonlinear effect on $y_i$. We assume that the outcome $y_i$ depends on features $\bx_i$ through the data generating model
\begin{align*}
y_i =\mu + h(\bx_i)+\epsilon_i, \quad \mbox{where}\; \epsilon_i \; \overset{iid}{\sim} N(0, \lambda). 
%\label{data_gen}
\end{align*}
We assume $h: \real^p \rightarrow \real$ follows the Gaussian process (GP) prior $\Gsc \Psc (0, k)$ governed by the positive definite kernel function $k$, such that the function evaluated at the observed covariates follows the multivariate normal (MVN) distribution
\begin{align*}
\bh =[h(\bx_1), \dots, h(\bx_n)] \sim MVN(\bzero, \bK)
\end{align*}
with covariance matrix having elements $\bK_{ij}=k(\bx_i, \bx_j)$. Under this construction, the predictive distribution of $h$ evaluated at the samples is also multivariate normal,
\begin{align*}
\bh \mid \{y_i, \bx_i\}_{i=1}^n &\sim MVN(\bh^*, \bK^*),\\
\bh^* &= \bK(\bK + \lambda \bI)^{-1}(\by-\bmu),\\
\bK^* &=\bK-\bK(\bK + \lambda \bI)^{-1}\bK.
\end{align*}
To understand the impact of $\lambda$ and $k$ on $\bh^*$, recall that operationally, a Gaussian process can be understood as the Bayesian version of kernel machine regression, where $\bh^*$ equivalently arises from the optimization problem
\begin{align*}
\bh^* =\underset{h \in \Hsc_k}{\mbox{argmin}} \; \lVert \by-\bmu-h(\bx) \rVert^2 + \lambda \lVert h \rVert_\Hsc^2,
\end{align*}
where $\Hsc_k$ is the reproducing kernel Hilbert space (RKHS) generated by kernel function $k$. From this perspective, $\bh^*$ is the element in a spherical ball in $\Hsc_k$ that best approximates the observed data $\by$. The mathematical properties (e.g., smoothness, spectral density) of $\bh^*$ are governed by the kernel function $k$. The norm of $\bh^*$, $\lVert h \rVert_\Hsc^2$, is constrained by the tuning parameter $\lambda$. 

Consequently, choice of kernel function $k$ and tuning parameter $\lambda$ critically impact the quality of the final estimate $\widehat{\bh}$. To this end, \pkg{CVEK} offers a wide range of choices for the kernel family and tuning parameter selection strategies, which we review below.

\subsubsection{Kernel Family}
In this section we review some commonly-used kernel functions that are implemented in \pkg{CVEK}, including three stationary covariance functions (Gaussian radial basis function, Mat\'{e}rn and rational quadratic), as well as non-stationary covariance functions (polynomial and neural network).
\begin{itemize}
\item \textbf{Intercept Kernel} (\code{intercept})

The intercept kernel implements the simplest of all kernel functions
\begin{align*}
k(\bx, \bx')=1, 
%\label{int_kern}
\end{align*}
which corresponds to the intercept under the (generalized) linear model.

\item \textbf{Linear Kernel} (\code{linear})

The linear kernel is
\begin{align*}
k(\bx, \bx')=\la \bx, \bx' \ra, 
%\label{lin_kern}
\end{align*}
where $\la \bx, \bx' \ra := \bx^\top \bx'$, the inner product of $\bx$ and $\bx'$. It is useful when dealing with large, sparse data vectors $\bx$.

\item \textbf{Polynomial Kernel} (\code{polynomial})

The polynomial kernel is
\begin{align*}
k(\bx, \bx')=(1+\la \bx, \bx' \ra)^p, 
%\label{poly_kern}
\end{align*}
which is commonly used with support vector machines (SVMs). The polynomial kernel becomes the intercept kernel when $p=0$, and the linear kernel when $p=1$.

\item \textbf{Gaussian Radial Basis Function (RBF) Kernel} (\code{rbf})

The Gaussian radial basis function kernel is
\begin{align*}
k(\bx, \bx')=\exp \Big(-\frac{\lvert \bx-\bx' \rvert^2}{2l^2}\Big), 
%\label{rbf_kern}
\end{align*}
where $l$ is the \emph{characteristic length-scale}. It is typically used when knowledge about the form of the exposure-response relationship exists.

\item \textbf{Mat\'{e}rn Kernel} (\code{matern})

The Mat\'{e}rn kernel is
\begin{align*}
k(\bx, \bx')=\frac{2^{1-\nu}}{\Gamma(\nu)}\Big(\frac{\sqrt{2\nu \lvert \bx-\bx' \rvert}}{l}\Big)^\nu K_\nu \Big(\frac{\sqrt{2\nu \lvert \bx-\bx' \rvert}}{l}\Big)
\end{align*}
with positive parameters $\nu$ and $l$, where $K_\nu$ is a modified Bessel function \citep{abramowitz_handbook_1974}. The Mat\'{e}rn kernel is commonly used to define the statistical covariance between measurements made at two points that are $\lvert \bx-\bx' \rvert$ units distant from each other. The most interesting cases for machine learning are $\nu = 3/2$ and $\nu = 5/2$, for which 
\begin{align*}
k_{\nu = 3/2}(\bx, \bx')&=\Big(1+\frac{\sqrt{3}\lvert \bx-\bx' \rvert}{l}\Big) \exp\Big(-\frac{\sqrt{3}\lvert \bx-\bx' \rvert}{l}\Big),\\
k_{\nu = 5/2}(\bx, \bx')&=\Big(1+\frac{\sqrt{5}\lvert \bx-\bx' \rvert}{l}+\frac{5\lvert \bx-\bx' \rvert^2}{3l^2}\Big) \exp\Big(-\frac{\sqrt{5}\lvert \bx-\bx' \rvert}{l}\Big),
\end{align*}
since for $\nu = 1/2$ the process becomes very rough, and for $\nu \geq 7/2$, in the absence of explicit prior knowledge about the existence of higher order derivatives, it is probably very hard from finite noisy training examples to distinguish between values of $\nu \geq 7/2$.

\item \textbf{Rational Quadratic Kernel} (\code{rational})

The rational quadratic kernel is
\begin{align*}
k(\bx, \bx')=\Big(1+\frac{\lvert \bx-\bx' \rvert^2}{2\alpha l^2}\Big)^{-\alpha} %\label{rq_kern}
\end{align*}
with $\alpha$, $l>0$ can be seen as a \emph{scale mixture} (an infinite sum) of squared exponential (SE) covariance functions with different characteristic length-scales (sum of covariance functions is a valid covariance). The limit of the rational quadratic covariance as $\alpha \rightarrow \infty$ is the SE covariance function with characteristic length-scale $l$.

\item \textbf{Neural Network Kernel} (\code{nn})

The neural network kernel is
\begin{align*}
k(\bx, \bx')=\frac{2}{\pi}\mbox{sin}^{-1}\Big(\frac{2\sigma \tilde{\bx}^\top \tilde{\bx}'}{\sqrt{(1+2\sigma \tilde{\bx}^\top \tilde{\bx})(1+2\sigma \tilde{\bx}'^\top \tilde{\bx}')}}\Big),
\end{align*}
where $\tilde{\bx}=(1, x_1, ..., x_d)^\top $ is an augmented input vector and $\sigma$ is the covariance coefficient.
\end{itemize}

\subsubsection{Model Selection Criteria} 
\label{sec:tuning}
In practice, the tuning parameter $\lambda$ is selected by minimizing certain  objective functions in a process known as \textit{Model Selection} that measures the model's degree of "appropriateness" given certain values of $\lambda$. Depending on the specific criteria, such "appropriateness" can be the distance between the current model and the true model, the model's out-of-sample prediction error, or the model likelihood. Here we review some commonly used model selection criteria.

In kernel machine regression, most of the model selection criteria can be expressed as a function of $\lambda$ through the model's predictive "hat" matrix $\bA_{\lambda}$:
\begin{align*}
\bA_\lambda=\bK(\bX, \bX)[\bK(\bX, \bX)+\lambda \bI]^{-1}.
\end{align*}
In this way, $\tr(\bA_\lambda)$ is the effective number of model parameters. It decreases monotonically with $\lambda>0$. For notational simplicity we assume $\by$ is centered: $\by=\by-\widehat{\bmu}$, where $\hat{\mu}=\frac{1}{n}\sum_{i=1}^ny_i$.
\begin{itemize}
\item \textbf{Akaike Information Criterion} (\code{AIC}) and its small-sample variant (\code{AICc})

AIC handles the trade-off between the goodness of fit of the model and the simplicity of the model:
\begin{align*}
\lambda_{AIC}=\underset{\lambda \in \Lambda}{\mbox{argmin}}\Big\{\log\; \by^{\top}(\bI-\bA_\lambda)^2\by+\frac{2[\tr(\bA_\lambda)+2]}{n}\Big\}, %\label{AIC}
\end{align*}

where $\Lambda$ is the set that contains all possible values of $\lambda$. When $n$ is small (e.g., $n/p < 40$) \citep{burnham_model_2002}, extreme overfitting is possible, giving small bias/ large variance estimates. The AIC small-sample correction \citep{hurvich_regression_1989, hurvich_clifford_m_smoothing_2002} is derived by modifying the penalty as the product of the original penalty, $2[\tr(\bA_\lambda)+2]/n$ and $n/[n-\tr(\bA_\lambda)-3]$, where we plug in $\bA_\lambda$ and $\hat{\sigma}^2$. In this case, we obtain our small-sample objective function AICc,
\begin{align*}
\lambda_{AICc}=\underset{\lambda \in \Lambda}{\mbox{argmin}}\Big\{\log\; \by^{\top}(\bI-\bA_\lambda)^2\by+\frac{2[\tr(\bA_\lambda)+2]}{n-\tr(\bA_\lambda)-3}\Big\}. 
%\label{AICc}
\end{align*}

\item \textbf{Bayesian Information Criterion} (\code{BIC})

The Bayesian information criterion is
\begin{align*}
\lambda_{BIC}=\underset{\lambda \in \Lambda}{\mbox{argmin}}\Big\{\log\; \by^{\top}(\bI-\bA_\lambda)^2\by+\frac{\log(n)[\tr(\bA_\lambda)+2]}{n}\Big\}. %\label{BIC}
\end{align*}
The formula for BIC is similar to the one for AIC. It is more conservative in the selection process with penalty for number of parameters to be $\log (n)$, instead of $2$ for AIC.

\item \textbf{Leave-one-out Cross Validation} (\code{loocv})

Suppose we perform $K$-fold cross-validation, which partitions observations into $K$ groups, $\kappa(1),...,\kappa(K)$, and calculates $\bA_\lambda$ $K$ times, each time leaving out group $\kappa(i)$, to get 
\[\bA_\lambda^{-\kappa(1)}, \bA_\lambda^{-\kappa(2)}, \dots, \bA_\lambda^{-\kappa(K)}.\]  
A value of $K = 10$ is very common in the field of applied machine learning. For $\bA_\lambda^{-\kappa(i)}$,  cross-validated residuals are calculated on the observations in $\kappa(i)$, which did not contribute to the estimation of $\bA$. The objective function estimates prediction error and is the sum of the squared cross-validated residuals,
\begin{align*}
\lambda_{K-CV}=\underset{\lambda \in \Lambda}{\mbox{argmin}}\;\Big\{\log \sum_{i=1}^K[\by_{\kappa(i)}-\bA_\lambda^{-\kappa(i)}\by_{\kappa(i)}]^\top[\by_{\kappa(i)}-\bA_\lambda^{-\kappa(i)}\by_{\kappa(i)}]\Big\}.
\end{align*}

Note that loocv corresponds to $K=n$. In this case, we can write our objective function as \citep{golub_generalized_1979},
\begin{align}\label{loocv}
\lambda_{loocv}=\underset{\lambda \in \Lambda}{\mbox{argmin}}\;\Big\{\log\;\by^{\top}[\bI-\diag(\bA_\lambda)-\frac{1}{n}\bI]^{-1}(\bI-\bA_\lambda)^2[\bI-\diag(\bA_\lambda)-\frac{1}{n}\bI]^{-1}\by \Big\}. 
%\label{loocv}
\end{align}

\item \textbf{Generalized Cross Validation} (\code{GCV}) and its small-sample variant (\code{GCVc})\\
In \eqref{loocv}, if we approximate each diagonal element of the matrix $\bA_\lambda$, $A_{\lambda[ii]}$  with its mean $\frac{\mbox{tr}(\bA_\lambda)}{n}$, in a sense we give equal weight to all observations. We then get the generalized cross validation objective function,
\begin{align}
\lambda_{GCV}=\underset{\lambda \in \Lambda}{\mbox{argmin}}\Big\{\log\; \by^{\top}(\bI-\bA_\lambda)^2\by-2\log[1-\frac{\tr(\bA_\lambda)}{n}-\frac{1}{n}]\Big\},
\label{GCV}
\end{align}

where '$-\frac{1}{n}$' is due to GCV counting $\mu$ as part of the model complexity, but not $\sigma^2$. This motivates the proposed small-sample correction to GCV \citep{boonstra_small-sample_2015}, which does count $\sigma^2$ as a parameter,
\begin{align*}
\lambda_{GCVc}=\underset{\lambda \in \Lambda}{\mbox{argmin}}\Big\{\log\; \by^{\top}(\bI-\bA_\lambda)^2\by-2\log[1-\frac{\tr(\bA_\lambda)}{n}-\frac{2}{n}]_+\Big\}.
\end{align*}

\item \textbf{Generalized Maximum Profile Marginal Likelihood} (\code{gmpml})

The generalized maximum profile marginal likelihood is defined as
\begin{align*}
\lambda_{gmpml}=\underset{\lambda \in \Lambda}{\mbox{argmin}}\Big\{\log\; \by^{\top}(\bI-\bA_\lambda)\by-\frac{1}{n-1}\log \lvert \bI-\bA_\lambda \rvert \Big\}. 
\end{align*}
This is a likelihood-based method, where $\lambda$ is interpreted as the variance component of a mixed-effects model.
\end{itemize}
%% -----------------------------------------------------------------------------

\subsection{Cross-Validated ensemble of kernels}
Traditional applications of Gaussian process estimate $h$ using a single kernel function $k$ for $h \in \Hsc_k$, therefore imposing \emph{a priori} assumption on the mathematical properties of $h$ through $k$. In such case, choosing a kernel function that is too restrictive or too flexible will lead to either model underfit or overfit, rendering the subsequent hypothesis tests invalid. Recently, \cite{liu_robust_2017, liu_cross-validated_2019} addressed the challenge by proposing Cross-Validated Ensemble of Kernels (CVEK), an ensemble-based estimator that adaptively learns the form of the kernel function from data. CVEK estimates $h$ using the ensemble of GP predictions generated from a library of (fixed) base kernel functions $\{k_d\}_{d=1}^D$,
\begin{align}
\label{eq:ens_h}
\hat{h}(\bx)=\sum_{d=1}^D u_d\hat{h}_d(\bx), \quad \bu \in \Delta :=\{\bu \mid \bu \geq 0, \parallel \bu \parallel_1=1\},
\end{align}
where $\hat{h}_d$ is the kernel predictor generated by the $d^{th}$ base kernel $k_d$.

The exact algorithm proceeds in three stages as follows (see Algorithm 1, which can be found in Section \ref{sec:append}).

\begin{enumerate}
\item[] \textbf{Stage 1: Estimate Base Model Cross-Validation Error}

For each base kernel in the library $\{ k_d \}_{d=1}^D$, we first standardize the kernel matrix by its trace $\bK_d = \bK_d/\tr(\bK_d)$, and then estimate the prediction based on each kernel as $\widehat{\bh}_{d, \hat{\lambda}_d} = \bK_d (\bK_d + \hat{\lambda}_d \bI)^{-1} \by, d\in\{1, \dots, D\}$, where the tuning parameter $\hat{\lambda}_d$ is selected by minimizing one of the model selection criterion introduced in Section \ref{sec:tuning}. In the case of leave-one-out cross validation (loocv), the cross-validation error can be expressed in closed-form: 
\begin{align*}
\texttt{CV}(\lambda | k_d) &= \left[ \bI - \diag(\bA_{d, \lambda})\right]^{-1}(\by - \widehat{\bh}_{d, \lambda}),  
\quad \mbox{where} \quad
\bA_{d, \lambda} = \bK_d (\bK_d + \lambda \bI)^{-1}.
\end{align*}
We denote the final estimated loocv error for $d^{th}$ kernel as $\hat{\xi}_d=\texttt{CV}(\hat{\lambda}_d | k_d)$

\item[]\textbf{Stage 2: Estimate Ensemble}

Using the estimated individual model cross-validation errors $\{\hat{\xi}_d\}_{d=1}^D$, we estimate the ensemble weights $\bu = \{u_d\}_{d=1}^D$ according to one of the ensemble strategies that will be introduced in Section \ref{sec:ensemble}.
After estimating $\widehat{\bu}$, the final ensemble prediction is estimated as:
$$
\widehat{\bh}
= \sum_{d=1}^D \hat{u}_d \widehat{\bh}_d 
= \sum_{d=1}^D \hat{u}_d  \bA_{d, \hat{\lambda}_d} \by 
= \widehat{\bA} \by,
$$
where $\widehat{\bA} = \sum_{d=1}^D \hat{u}_d \bA_{d, \hat{\lambda}_d}$ is the ensemble hat matrix.

\item[] \textbf{Stage 3: Estimate Ensemble Kernel Matrix} \\
Using the ensemble hat matrix $\widehat{\bA}$, estimate the ensemble kernel matrix $\widehat{\bK}$ by solving:
\begin{align*}
\widehat{\bK} (\widehat{\bK} + \lambda \bI)^{-1} = \widehat{\bA}.
\end{align*}
Specifically, if we denote $\bU_A$ and $\{\delta_{A,k}\}_{k=1}^n$ as the eigenvectors and eigenvalues of $\widehat{\bA}$, respectively, then the ensemble kernel matrix $\widehat{\bK}$ adopts the form:
\begin{align}
\label{eq:K_ens}
\widehat{\bK} = \lambda_\bK * \bigg[
\bU_A \diag \Big( \frac{\delta_{A, k}}{1 - \delta_{A,k}} \Big) \bU_A^\top \bigg],
\end{align}
where we recommend setting $\lambda_\bK = \mbox{min}\bigg[1,  \Big( \sum_{k=1}^n \frac{\delta_{A, k}}{1 - \delta_{A,k}} \Big)^{-1} \bigg]$.
\end{enumerate}

\subsubsection{Ensemble Strategies}
We now introduce the choices for an ensemble strategy to be used  in Stage 2 of CVEK. Briefly, an ensemble strategy estimates the ensemble weights $\bu=\{u_d\}_{d=1}^D$ from individual model cross-validation errors $\{\hat{\xi}_d\}_{d=1}^D$. Choices available are:
\label{sec:ensemble}
\begin{itemize}
\item \textbf{Averaging Ensemble} (\code{avg}) \\
Motivated by existing literature on the omnibus kernel \citep{zhan_fast_2017}, a simple way to choose the weights is $u_d=1/D$ for $d=1,2,...D$.

\item \textbf{Exponential Weighting} (\code{exp}) \\
\cite{dalalyan_aggregation_2007} proposed estimating $\bu$ using the estimated errors $\{\hat{\xi}_d\}_{d=1}^D$ as:
\begin{align*}
u_d(\beta)=\frac{\exp(-\lVert \hat{\xi}_d \rVert_2^2/\beta)}{\sum_{d=1}^D \exp(-\lVert \hat{\xi}_d \rVert_2^2/\beta)}.
\end{align*}
From the perspective of optimal model aggregation, the authors showed that under squared loss, the error bound in excessive risk of exponential weighting converges at the fast rate of $O(\frac{1}{n})$. Here, $\beta$ serves as a tuning parameter to determine how different the ensemble weights are. Increasing $\beta$ results in more similar weights, based on their estimated errors. An infinite $\beta$ results in an averaging ensemble.  Exponential weighting can also be viewed as the frequentist version for Bayesian model averaging under Gaussian noise \citep{yang_minimax_2014}.

\item \textbf{Cross-Validated Stacking} (\code{stack}) \\
Alternatively, we can estimate $\bu$ such that it minimizes the overall cross-validation error. After obtaining the estimated errors $\{\hat{\xi}_d\}_{d=1}^D$, we estimate the ensemble weights $\bu=\{u_d\}_{d=1}^D$ such that it minimizes the overall error:
\begin{align*}
\hat{\bu}=\underset{\bu \in \Delta}{\mbox{argmin}}\lVert \sum_{d=1}^Du_d\hat{\xi}_d\rVert^2 \quad \mbox{where}\; \Delta=\{\bu \mid \bu \geq 0, \lVert \bu \rVert_1=1\}.
\end{align*}
\end{itemize}

\raggedright
\section{Hypothesis testing for nonlinear effects}
\label{sec:test}
\subsection{Testing for general nonlinear effect}
We use the classical variance component test \citep{lin_variance_1997} to construct a testing procedure for the hypothesis about a Gaussian process. Specifically, recall the assumed model:
\begin{align*}
y_i =\mu + h(\bx_i)+\epsilon_i, \quad \mbox{where}\; \epsilon_i \; \overset{iid}{\sim} N(0, \lambda). 
\end{align*}
We are interested in testing the null hypothesis:
\begin{align*}
H_0: h \in \Hsc_0.
\end{align*}
We first express this hypothesis in terms of model parameters. The key to our approach is to assume that $h$ lies in a RKHS generated by a \emph{garrote kernel function} $k_{\delta}(\bz, \bz')$ \citep{maity_powerful_2011}, which is constructed by including an extra \emph{garrote parameter} $\delta$ in a given kernel function. When $\delta=0$, the garrote kernel function $k_0(\bx, \bx')=k_{\delta}(\bx, \bx')\mid _{\delta=0}$ generates $\Hsc_0$, the space of functions under the null hypothesis. In order to focus on a particular hypothesis of interest, practitioners need only to specify the form of the garrote kernel such that $\Hsc_0$ corresponds to the null hypothesis. As a result, the general null hypothesis is equivalent to,
\begin{align}
H_0: \delta=0. \label{hypo}
\end{align}
We now construct a test statistic $\hat{T}_0$ for \eqref{hypo} by noticing that the garrote parameter $\delta$ can be treated as a variance component parameter in a linear mixed model (LMM). This is because the Gaussian process under a garrote kernel can be formulated into the LMM
\begin{align*}
\by=\bmu+\bh+\bepsilon, \quad \mbox{where} \quad \bh \sim N(\mathbf{0}, \tau \bK_\delta), \quad \bepsilon \sim N(\mathbf{0}, \sigma^2\bI),
\end{align*}
and $\bK_\delta$ is the kernel matrix generated by $k_{\delta}(\bz, \bz')$. Consequently, we can derive a variance component test for $H_0$ by calculating the squared derivative of the Restricted Maximum Likelihood (REML) with respect to $\delta$ under $H_0$ \citep{lin_variance_1997},
\begin{align}
\hat{T}_0=\hat{\tau}*(\by-\hat{\bmu})^\top \bV_0^{-1}[\partial \bK_0]\bV_0^{-1}(\by-\hat{\bmu}), \label{test_stat}
\end{align}
where $\tau = \frac{\sigma^2}{\lambda}$ and  $\bV_0=\hat{\sigma}^2\bI+\hat{\tau}\bK_0$. In this expression, $\bK_0=\bK_\delta \mid_{\delta=0}$, and $\partial \bK_0$ is the null derivative kernel matrix whose $(i, j)^{th}$ entry is $\frac{\partial }{\partial \delta}k_\delta(\bx, \bx') \mid_{\delta=0}$. Appendix \ref{appendix_E} also provides the derivation of the REML based test statistic.

\subsubsection{Extension for Interaction Testing}

In the previous section, we assume that we are able to obtain a $k_\delta$ that generates $\Hsc_0$ exactly. However, depending on the exact hypothesis of interest, identifying such a $k_0$ is not always straightforward. In this section, we revisit the case of interaction testing and consider how to build a $k_0$ for the hypothesis of interest.
\begin{align*}
\Hsc_0: &\; h(\bx)=h_1(\bx_1)+h_2(\bx_2), \\
\Hsc_a: &\; h(\bx)=h_1(\bx_1)+h_2(\bx_2)+h_{12}(\bx_1, \bx_2),
\end{align*}
where $h_{12}$ is the ``pure interaction'' function that is orthogonal to main effect function $h_1$ and $h_2$. This hypothesis is difficult to formulate with Gaussian process models, since the kernel functions $k(\bx, \bx')$ in general do not explicitly separate the main and the interaction effect. Therefore rather than directly defining $k_0$, we need to first construct $\Hsc_0$ and $\Hsc_a$ that correspond to the null and alternative hypotheses, respectively, and then identify the garrote kernel function $k_\delta$ such that corresponds to $\Hsc_0$ when $\delta=0$ and $\Hsc_a$ when $\delta >0$.

We build $\Hsc_0$ using the tensor-product constructions of RKHS on the product domain\\ $(\bx_{1, i}, \bx_{2, i}) \in \real^{p_1} \times \real^{p_2}$ \citep{gu_smoothing_2013}, due to this approach's unique ability to explicitly characterize the space of ``pure interaction'' functions. Let $\bone =\{f \mid f \propto 1\}$ be the RKHS of constant functions, and $\Hsc_1$, $\Hsc_2$ be the RKHS of centered functions for $\bx_1$, $\bx_2$ respectively. We can then define the full space as $\Hsc=\otimes_{m=1}^2 (\bone \oplus \Hsc_m)$. $\Hsc$ describes the space of functions that depends jointly on $\{\bx_1, \bx_2\}$. It adopts the orthogonal decomposition,
\begin{align*}
\Hsc &=(\bone \oplus \Hsc_1)\otimes (\bone \oplus \Hsc_2)\\
&=\bone \otimes \{\Hsc_1 \oplus \Hsc_2\} \oplus \{\Hsc_1 \otimes \Hsc_2\}=\bone \oplus \Hsc_{12}^\perp \oplus \Hsc_{12}, 
\end{align*}
where we have denoted $\Hsc_{12}^\perp=\Hsc_1 \oplus \Hsc_2$ and $\Hsc_{12}=\Hsc_1 \otimes \Hsc_2$ respectively. We see that $\Hsc_{12}$ is indeed the space of ``pure interaction'' functions, since $\Hsc_{12}$ contains functions on the product domain $\real^{p_1} \times \real^{p_2}$, but is orthogonal to the space of additive main effect functions $\Hsc_{12}^\perp$. To summarize, we have identified two function spaces $\Hsc_0$ and $\Hsc_a$ that have the desired interpretation,
\begin{align*}
\Hsc_0=\Hsc_{12}^\perp, \quad \Hsc_a=\Hsc_{12}^\perp \oplus \Hsc_{12}.
\end{align*}
We are now ready to identify the garrote kernel $k_\delta(\bx, \bx')$. To this end, we notice that both $\Hsc_{12}^\perp$ and $\Hsc_{12}$ are composite spaces built from basis RKHSs using direct sum and tensor products. If we denote $k_m(\bx_m, \bx'_m)$ as the reproducing kernel associated with $\Hsc_m$, we can construct kernel functions for composite spaces $\Hsc_{12}^\perp$ and $\Hsc_{12}$ as 
\begin{align*}
k_0(\bx, \bx') &=k_1(\bx_1, \bx_1)+k_2(\bx_2, \bx_2),\\
k_{12}(\bx, \bx') &=k_1(\bx_1, \bx_1)k_2(\bx_2, \bx_2).
\end{align*}
Hence, the garrote kernel function for $\Hsc_a$ is
\begin{align*}
k_\delta(\bx, \bx') &=k_0(\bx, \bx')+\delta k_{12}(\bx, \bx').
\end{align*}
Finally, using the chosen form of the garrote kernel function, the $(i, j)^{th}$ element of the null derivative kernel matrix is $\frac{\partial }{\partial \delta}k_\delta(\bx, \bx') \Big|_{\delta=0} =k_{12}(\bx, \bx')$, i.e., the null derivative kernel matrix $\partial \bK_0$ is simply the kernel matrix $\bK_{12}$ that corresponds to the interaction space. Thus, we also call it the alternative kernel. Therefore the score test statistic $\hat{T}_0$ in $\eqref{test_stat}$ simplifies to
\begin{align*}
\hat{T}_0=\hat{\tau}*(\by-\bX \hat{\bbeta})^\top \bV_0^{-1}\bK_{12}\bV_0^{-1}(\by-\bX \hat{\bbeta}),
\end{align*}
where $\bV_0=\hat{\sigma}^2 \bI + \hat{\tau} \bK_0$. \\

There exist multiple approaches for estimating the null distribution of $\hat{T}_0$. Specifically, assuming $\partial \bK_{0}$ is a fixed matrix (e.g., in the case of interaction testing, fixing the $k_{12}$ to be a linear kernel), one can derive the closed form expression of the asymptotic distribution of $\hat{T}_0$ (i.e., a mixture of $\chi^2$ distributions). Alternatively, one can approximate the null distribution in a data-driven manner using bootstrap sampling. Compared to the bootstrap, the asymptotic method is advantageous in that it is more powerful if the null model is correctly specified, but is restrictive in that it requires the alternative
kernel $\partial \bK_{0}$ to be fixed \textit{a priori}. This requirement further prevents practitioners from improving test power due to the need to learn the optimal $\partial \bK_{0}$ from data. The bootstrap test, on the other hand, does not require $\partial \bK_{0}$ to be fixed and hence does not suffer from this limitation. We introduce these two types of procedures in detail in Section \ref{sec:null_dist}, and discuss a data-adaptive strategy for estimating $\partial \bK_{0}$ in Section \ref{sec:alt_kernel}.

\subsection{Null distribution estimation}
\label{sec:null_dist}
\subsubsection{Asymptotic Approximation}
Assuming fixed $\partial \bK$, the null distribution of $\hat{T}$ can be approximated with a scaled chi-square distribution $\kappa \chi_\nu^2$ using the Satterthwaite method that matches the first two moments of $T$,
\begin{align*}
\kappa * \nu=\E(T)=\hat{\tau} * \tr(\bV_0\partial \bK_0),
 \quad 2*\kappa^2*\nu=Var(T)=\hat{\bI}_{\delta\delta}. 
\end{align*}
This procedure yields the solution
\begin{align*}
\hat{\kappa}=\hat{\bI}_{\delta\delta}/[\hat{\tau}*\tr(\bV_0^{-1}\partial \bK_0)], \quad \hat{\nu}=[\hat{\tau}*\tr(\bV_0^{-1}\partial \bK_0)]^2/(2*\hat{\bI}_{\delta\theta}),
\end{align*}
where $\hat{\bI}_{\delta\delta}=\bI_{n, \delta\delta}-\bI_{\delta\theta}^\top \bI_{\theta\theta}^{-1}\bI_{\delta\theta}$ is the efficient information of $\delta$ REML. $\bI_{\delta\delta}$, $\bI_{\theta\theta}$ and $\bI_{\delta\theta}$ are sub-matrices of the REML information matrix. 
Numerically more accurate, but computationally less efficient, approximation methods are also available \citep{bodenham_comparison_2016}.

Finally, the p-value of this test is calculated using the tail probability of $\hat{\kappa} \chi_{\hat{\nu}}^2$,
\begin{align*}
p=P(\hat{\kappa} \chi_{\hat{\nu}}^2>\hat{T})=P(\chi_{\hat{\nu}}^2>\hat{T}/\hat{\kappa}).
\end{align*}

A complete summary of the proposed testing procedure is available in Algorithm 2, which can be found in Section \ref{sec:append}.

\subsubsection{Parametric Bootstrap}
When the sample size is small, we make valid inferences about a population using resampling. A commonly used resampling method is the bootstrap, which gives valid tests in small to moderate sample sizes.

Testing in a regression model framework requires computing the distribution of the test statistic under the null hypothesis. We approximate this null distribution using a  bootstrap sample of the test statistic resampled from the fit of the null model. For instance, when testing \eqref{hypo}, we first fit the model under the null,
\begin{align*}
\E(\by^\star)=\bK_0(\bK_0+\lambda \bI)^{-1}\by=\bA_0\by,
\end{align*}
and generate $\bY^\star$ with a random noise, whose variance is also estimated. We then compute the test statistic for this simulated sample, and repeat this process $B$ times. The empirical distribution of the test statistic provides an estimate of the test statistic's distribution under the null. Correspondingly, p-values are calculated as the proportion of simulated test statistics that are as or more extreme than the observed value.

Like the classical bootstrap, this approach samples from a distribution based on the observed data, but the simulations are from a fitted parametric model rather than the empirical distribution. To obtain a valid test, the fitted parametric model is chosen so that the null hypothesis is satisfied.
A complete summary of the proposed testing procedure is available in Algorithm 3, which can be found in Section \ref{sec:append}.

\subsection{Strategy for estimating alternative kernel}
\label{sec:alt_kernel}
As mentioned previously, the asymptotic test requires the alternative
kernel $\partial \bK_0$ in the test statistic (\ref{test_stat}) to be fixed \textit{a priori}, due to the need to approximate the null distribution analytically. Traditionally, $\partial \bK_0$ is fixed to be either linear or a specific kernel family with fixed hyperparameters (e.g., Gaussian RBF family with fixed length-scale). Consequently, the form of the alternative kernel needs to be  correctly specified in order to sufficiently describe the interaction effect, since otherwise a misspecified alternative
kernel may lead to a loss of power. On the other hand, the bootstrap test allows $\partial \bK_0$ to be estimated adaptively from the data in order to better represent the alternative hypothesis space.

To this end, we propose a strategy for data-adaptive estimation of the alternative kernel in the bootstrap test. Specifically, we estimate the alternative kernel using the ensemble weights $\{\hat{\mu}_d\}_{d=1}^D$ obtained from the ensemble procedure as described in \eqref{eq:ens_h}, i.e.,
\begin{align}
\label{eq:ens_alt}
\partial \bK_0=\sum_{d=1}^D \hat{u}_d * \partial \bK_{0, d}. 
\end{align}
Consequently, if the true interaction effect is not linear,  a bootstrap test with an adaptively estimated alternative kernel can better describe the interaction effect from the data, and therefore will have better power when compared to an asymptotic or bootstrap test with the $\partial \bK_0$ fixed to be a linear kernel. In Section \ref{sec:simu_test_type}, we empirically investigate the effectiveness of this approach through extensive simulation.

%% -- Illustrations ------------------------------------------------------------

%% - Virtually all JSS manuscripts list source code along with the generated
%%   output. The style files provide dedicated environments for this.
%% - In R, the environments {Sinput} and {Soutput} - as produced by Sweave() or
%%   or knitr using the render_sweave() hook - are used (without the need to
%%   load Sweave.sty).
%% - Equivalently, {CodeInput} and {CodeOutput} can be used.
%% - The code input should use "the usual" command prompt in the respective
%%   software system.
%% - For R code, the prompt "R> " should be used with "+  " as the
%%   continuation prompt.
%% - Comments within the code chunks should be avoided - these should be made
%%   within the regular LaTeX text.
\section{The CVEK package} 
\label{sec:package}

Using a library of base kernels, \pkg{CVEK} learns the  generating function from data by directly minimizing the ensemble model's error, and tests whether the data is generated by the RKHS under the null hypothesis. Section \ref{tutor:simu} presents a simple example to conduct Gaussian process regression and hypothesis testing using the \code{cvek} function on simulated data. Section \ref{tutor:real} shows a real-world application where we use \pkg{CVEK} to understand whether the per capita crime rate impacts the relationship between local socioeconomic status and the housing price in Boston, MA, U.S.A.

\subsection{Tutorial using simulated dataset}
\label{tutor:simu}
\subsubsection{Generate Data and Define Model}
We generate a simulated dataset using the \code{linear} kernel, and set the relative interaction strength to be $0.2$. The outcome $y_i$ is generated as,
\begin{align*}
y_i=h_1(\bx_{i, 1})+h_2(\bx_{i, 2})+0.2 * h_{12}(\bx_{i, 1}, \bx_{i, 2})+\epsilon_i,
\end{align*}
where $h_1$, $h_2$, $h_{12}$ are sampled from RKHSs $\Hsc_1$, $\Hsc_2$, $\Hsc_{12}$, generated using the corresponding \code{linear} kernel. We standardize all sampled functions to have unit form, so that $0.2$ represents the strength of interaction relative to the main effect.

\begin{CodeChunk}
\begin{CodeInput}
> set.seed(0726)
> n <- 60 # including training and test observations
> d <- 4
> int_effect <- 0.2
> data <- matrix(rnorm(n * d), ncol = d)
> Z1 <- data[, 1:2]
> Z2 <- data[, 3:4]
> 
> kern <- generate_kernel(method = "linear")
> w <- rnorm(n)
> w12 <- rnorm(n)
> K1 <- kern(Z1, Z1)
> K2 <- kern(Z2, Z2)
> K1 <- K1 / sum(diag(K1)) # standardize kernel
> K2 <- K2 / sum(diag(K2))
> h0 <- K1 %*% w + K2 %*% w
> h0 <- h0 / sqrt(sum(h0 ^ 2)) # standardize main effect
> 
> h1_prime <- (K1 * K2) %*% w12 # interaction effect
> 
> # standardize sampled functions to have unit norm, so that 0.2
> # represents the interaction strength relative to main effect
> Ks <- svd(K1 + K2)
> len <- length(Ks$d[Ks$d / sum(Ks$d) > .001])
> U0 <- Ks$u[, 1:len]
> h1_prime_hat <- fitted(lm(h1_prime ~ U0))
> h1 <- h1_prime - h1_prime_hat
> 
> h1 <- h1 / sqrt(sum(h1 ^ 2)) # standardize interaction effect
> Y <- h0 + int_effect * h1 + rnorm(1) + rnorm(n, 0, 0.01)
> data <- as.data.frame(cbind(Y, Z1, Z2))
> colnames(data) <- c("y", paste0("z", 1:d))
> 
> data_train <- data[1:40, ]
> data_test <- data[41:60, ]
\end{CodeInput}
\end{CodeChunk}

The resulting data look as follows.
\begin{table}[h]
\begin{center}
\begin{tabular}{l|llll}
\textbf{y} & \textbf{z1} & \textbf{z2} & \textbf{z3} & \textbf{z4} \\ \hline
1.2065 & -0.354  & -0.8478 & -1.9983 & 1.3628  \\
1.5957 & -1.3522 & 0.9002  & 1.0221  & -0.7188 \\
1.4699 & 0.5276  & -0.8568 & 0.0372  & 0.4386  \\
1.5936 & -1.0577 & 0.7019  & 0.9086  & -0.8035 \\
1.3631 & 0.9927  & 0.7144  & -0.9476 & -0.2037  
\end{tabular}
\end{center}
\end{table}

Now we can apply the \code{cvek} function to conduct Gaussian process regression. Table 1 is a detailed list of all the arguments of the function \code{cvek}.
\begin{table}[h!]
\begin{center}
\begin{tabular}{ll} 
\hline
\textbf{Arguments} & \textbf{Description} \\ \hline
\code{formula} & (formula) A user-supplied formula for the null model. \\
\code{kern_func_list} & (list) A list of kernel functions in the model library. \\
\code{data} & (data.frame, n*d) A data.frame, list or environment (or object coercible \\
& by as.data.frame to a data.frame), containing the variables in formula. \\
& Neither a matrix nor an array will be accepted. \\
\code{formula_test} & (formula) A user-supplied formula indicating the alternative effect to \\
& test. All terms in the alternative mode must be specified as kernel terms. \\
\code{mode} & (character) A character string indicating which tuning parameter criteria \\
& is to be used. \\
\code{strategy} & (character) A character string indicating which ensemble strategy is to \\ 
& be used. \\
\code{beta_exp} & (numeric/character) A numeric value specifying the parameter when \\
& strategy = "exp". \\
\code{lambda} & (numeric) A numeric string specifying the range of tuning parameter to\\
& be chosen. The lower limit of lambda must be above 0. \\
\code{test} & (character) Type of hypothesis test to conduct. Must be either 'asymp' \\
& or 'boot'. \\
\code{alt_kernel_type} & (character) Type of alternative kernel effect to consider. Must be either \\
& 'linear' or 'ensemble'.\\
\code{B} & (numeric) Number of bootstrap samples. \\
\code{verbose} & (logical) Whether to print additional messages.\\ \hline
\end{tabular}
\caption{Arguments of the function \code{cvek()}.}
\end{center}
\end{table}

Suppose we want our model library to contain three kernels: \code{linear}, \code{polynomial} with \code{p=2}, and \code{rbf} with \code{l=1} (the effective parameter for \code{polynomial} is \code{p} and the effective parameter for \code{rbf} is \code{l}, so we can set anything to \code{l} for \code{polynomial} kernel and \code{p} for \code{rbf} kernel). We then first apply \code{define_library}.

\begin{CodeChunk}
\begin{CodeInput}
> kern_par <- data.frame(method = c("linear", "polynomial", "rbf"), 
+                        l = rep(1, 3), p = 1:3, stringsAsFactors = FALSE)
> # define model library
> kern_func_list <- define_library(kern_par)
\end{CodeInput}
\end{CodeChunk}

The null model is then $y \sim z1 + z2 + k(z3, z4)$.

\begin{CodeChunk}
\begin{CodeInput}
> formula <- y ~ z1 + z2 + k(z3, z4)
\end{CodeInput}
\end{CodeChunk}

\subsubsection{Estimation and Testing}

With all these parameters specified, we can conduct Gaussian process regression.

\begin{CodeChunk}
\begin{CodeInput}
> est_res <- cvek(formula, kern_func_list = kern_func_list, data = data_train)
> est_res$lambda
\end{CodeInput}
\begin{CodeOutput}
[1] 4.539993e-05
\end{CodeOutput}
\begin{CodeInput}
> est_res$u_hat
\end{CodeInput}
\begin{CodeOutput}
[1] 0.994864707 0.000000000 0.005135293
\end{CodeOutput}
\end{CodeChunk}

We can see that the ensemble weight assigns \code{0.99} to the \code{linear} kernel, which is the true kernel. This illustrates the accuracy and efficiency of the \pkg{CVEK} method.

We next specify the testing procedure. Note that we can use the same function \code{cvek} to perform hypothesis testing, as we did for estimation, but we need to provide \code{formula\_test}, which is the user-supplied formula indicating the additional alternative effect (e.g., interactions) to test for. Specifically, we will first show how to conduct the classic score test by specifying \code{test="asymp"}, followed by a bootstrap test where we  specify \code{test="boot"}, and the number of bootstrap samples \code{B=200}.

\begin{CodeChunk}
\begin{CodeInput}
> formula_test <- y ~ k(z1, z2):k(z3, z4)
> 
> cvek(formula, kern_func_list = kern_func_list, 
+      data = data_train, formula_test = formula_test,
+      mode = "loocv", strategy = "stack",
+      beta_exp = 1, lambda = exp(seq(-10, 5)),
+      test = "asymp", alt_kernel_type = "ensemble",
+      verbose = FALSE)$pvalue
\end{CodeInput}
\begin{CodeOutput}
             [,1]
[1,] 1.493613e-08
\end{CodeOutput}
\end{CodeChunk}

\begin{CodeChunk}
\begin{CodeInput}
> cvek(formula, kern_func_list = kern_func_list, 
+      data = data_train, formula_test = formula_test,
+      mode = "loocv", strategy = "stack",
+      beta_exp = 1, lambda = exp(seq(-10, 5)),
+      test = "boot", alt_kernel_type = "ensemble",
+      B = 200, verbose = FALSE)$pvalue
\end{CodeInput}
\begin{CodeOutput}
[1] 0
\end{CodeOutput}
\end{CodeChunk}

Both tests come to the same conclusion. At the significance level $0.05$, we reject the null hypothesis that there's no interaction effect, which matches our data generation mechanism. Additionally, we can predict new outcomes based on estimation results \code{est_res}.

\begin{CodeChunk}
\begin{CodeInput}
> y_pred <- predict(est_res, data_test[, 2:5])
> data_test_pred <- cbind(y_pred, data_test)
\end{CodeInput}
\end{CodeChunk}

\begin{table}[h]
\begin{center}
\begin{tabular}{l|l|llll}
\textbf{y\_pred} & \textbf{y} & \textbf{z1} & \textbf{z2} & \textbf{z3} & \textbf{z4} \\ \hline
1.4597  & 1.4552 & 0.4552  & 0.8838  & -0.1065 & -0.7295 \\
1.5226  & 1.5927 & -0.3551 & -0.7374 & -1.0941 & -2.1262 \\
1.4995  & 1.5197 & -2.9798 & -1.0641 & -0.2713 & -0.2268 \\
1.4939  & 1.5176 & 0.0146  & -0.2485 & -0.4155 & -0.9279 \\
1.487   & 1.5309 & -0.2603 & -1.0785 & -0.8478 & -0.8378   
\end{tabular}
\end{center}
\end{table}

\subsection{Detecting nonlinear interactions in Boston Housing Prices}
\label{tutor:real}
In this section, we show an example of using the \texttt{cvek} test to detect
nonlinear interactions between socioeconomic factors that contribute to housing price in the city of Boston,  Massachusetts, USA. We consider the \texttt{Boston} dataset (available in the \pkg{MASS} package), which is collected by the U.S Census Service about the median housing price (\code{medv}) in Boston, along with additional variables describing local socioeconomic information such as per capita crime rate, proportion of non-retail business, number of rooms per household, etc. Table 2 lists the $14$ variables. 

\begin{table}[h!]
\begin{center}
\begin{tabular}{ll} 
\hline
\textbf{Variables} & \textbf{Description} \\ \hline
\code{crim} & Per capita crime rate by town.\\
\code{zn} & Proportion of residential land zoned for lots over 25,000 sq.ft.\\
\code{indus} & Proportion of non-retail business acres per town.\\
\code{chas} & Charles River dummy variable (= 1 if tract bounds river; 0 otherwise).\\
\code{nox} & Nitric oxides concentration (parts per 10 million).\\
\code{rm} & Average number of rooms per dwelling. \\
\code{age} & Proportion of owner-occupied units built prior to 1940.\\
\code{dis} & Weighted distances to five Boston employment centres.\\
\code{rad} & Index of accessibility to radial highways.\\
\code{tax} & Full-value property-tax rate per USD 10,000.\\
\code{ptratio} & Pupil-teacher ratio by town.\\
\code{black} & $1000(B - 0.63)^2$ where B is the proportion of blacks by town.\\
\code{lstat} & Percentage of lower socioeconomic status in the population. \\
\code{medv} & Median value of owner-occupied homes in USD 1000's.\\ \hline
\end{tabular}
\caption{Variables of the \texttt{Boston} dataset.}
\end{center}
\end{table}

Here we use \texttt{cvek} to study whether the per capita crime rate (\texttt{crim}) impacts the relationship between the local socioeconomic status (\code{lstat}) and the housing price. The null model is,
\begin{align*}
medv \sim \mathbf{x}^\top \boldsymbol{\beta} + k(crim) + k(lstat),
\end{align*}
where $\mathbf{x}^\top=$\code{(1, zn, indus, chas, nox, rm, age, dis, rad, tax, ptratio, black)}, and $k()$ is specified as a semi-parametric model with a model library that includes \code{linear} and \code{rbf} kernels with $l=1$. This inclusion of nonlinearity (i.e., the \texttt{rbf} kernel) is important, since per classic results in the macroeconomics literature, the crime rates and socioeconomic status of local community are known to have a nonlinear association with the local housing price \citep{harrison_hedonic_1978}.
\begin{CodeChunk}
\begin{CodeInput}
> kern_par <- data.frame(method = c("linear", "rbf"), 
+                        l = rep(1, 2), p = 1:2, stringsAsFactors = FALSE)
> # define kernel library
> kern_func_list <- define_library(kern_par)
\end{CodeInput}
\end{CodeChunk}

To this end, the hypothesis regarding whether the crime rate (\code{crim}) impacts the association between local socioeconomic status (\code{lstat}) and the housing price (\code{medv}) is equivalent to testing whether there exists a nonlinear interaction between \texttt{crim} and \texttt{lstat} in predicting \texttt{medv}, i.e.,
\begin{align*}
\Hsc_0: &\; medv \sim \mathbf{x}^\top \boldsymbol{\beta} + k(crim) + k(lstat), \\
\Hsc_a: &\; medv \sim \mathbf{x}^\top \boldsymbol{\beta} + k(crim) + k(lstat) + k(crim):k(lstat).
\end{align*}

To test this hypothesis using \texttt{cvek}, we specify the null model using \texttt{formula}, and specify the additional interaction term ($k(crim):k(lstat)$) in the alternative model using \texttt{formula\_test}, as shown below:

\begin{CodeChunk}
\begin{CodeInput}
> formula <- medv ~ zn + indus + chas + nox + rm + age + dis + 
+   rad + tax + ptratio + black + k(crim) + k(lstat)
> formula_test <- medv ~ k(crim):k(lstat)
> fit_bos<- cvek(formula, kern_func_list = kern_func_list, data = Boston, 
+                formula_test = formula_test, 
+                lambda = exp(seq(-3, 5)), test = "asymp")
\end{CodeInput}
\end{CodeChunk}

Given the fitted object (\code{fit_bos}), the p-value of the \code{cvek} test can be extracted as below:
\begin{CodeChunk}
\begin{CodeInput}
> fit_bos$pvalue
\end{CodeInput}
\begin{CodeOutput}
[1] 4.614106e-06
\end{CodeOutput}
\end{CodeChunk}

Since $p<0.05$, we reject the null hypothesis that there's no \code{crim:lstat} interaction, and conclude that the data does suggest an impact of the crime rate on the relationship between the local socioeconomic status and the  housing price. In Appendix \ref{appendix_F}, we provide additional code showing how to visualize the interaction effect from a \code{cvek} model.

%% -- Summary/conclusions/discussion -------------------------------------------

\section{Simulation and practical recommendations}
\label{sec:simu}
In this section, we conduct a simulation study to evaluate the finite-sample performance of the CVEK hypothesis tests in a setting that is analogous to a typical nutrition-environment interaction study \citep{liu_cross-validated_2019}. We generate two groups of input features $(\bx_{i, 1}, \bx_{i,2}) \in \mathbb{R}^{p_1} \times \mathbb{R}^{p_2}$ independently, both within and between group from a standard Gaussian distribution, representing a subject's level of exposure to $p_1$ environmental pollutants and the levels of a subject's intake of $p_2$ nutrients. Across all simulation scenarios, we keep $n=100$, and $p_1=p_2=2$. We generate the outcome $y_i$ as,
\begin{align}
\label{eq:gen_data}
y_i=h_1(\bx_{i, 1})+h_2(\bx_{i, 2})+\delta * h_{12}(\bx_{i, 1}, \bx_{i, 2})+\epsilon_i,
\end{align}
where $h_1$, $h_2$, $h_{12}$ are sampled from RKHSs $\Hsc_1$, $\Hsc_2$, $\Hsc_{12}$, generated using a ground-truth main effect kernel $k_{main}$ (related to $h_1$, $h_2$) and an interaction kernel $k_{int}$ (related to $h_{12}$), and $\epsilon_i \sim N(0, \sigma^2=0.01^2)$. We standardize all sampled functions to have unit form, so that $\delta$ represents the strength of interaction relative to the main effect. Additional simulation results for correlated exposures are presented in Appendix \ref{appendix_B}.

For each simulation scenario, we first generate data using $\delta$, $k_{main}$ and $k_{int}$ as in (\ref{eq:gen_data}), then select a $k_{model}$ to estimate the null model and obtain a p-value using either an asymptotic approximation (Algorithm 2) or parametric bootstrap (Algorithm 3). We repeat each scenario 200 times, and evaluate the test performance using the empirical probability $\hat{P}(p\leq 0.05)$. Under the null hypothesis, a correct test should produce $\hat{P}$ that is smaller or equal to the significance level 0.05. Under the alternative hypothesis $H_a: \delta>0$, $\hat{P}(p\leq 0.05)$ estimates the test's power, and should ideally approach 1.0 quickly as the strength of interaction $\delta$ increases.

In this study, we vary the combination of $k_{main}$ (the true main effect) and $k_{int}$ (the true interaction effect) to produce data generating functions $h_\delta(\bx_{i, 1}, \bx_{i, 2})$ with different smoothness and complexity properties, and vary $k_{model}$ to reflect different common modeling strategies for the null model in addition to using CVEK. We then evaluate how these two aspects impact the Type I error and power of the resulting hypothesis tests.

Specifically, we consider the following seven types of data generation mechanism where $k_{main} = k_{int}$ (i.e., the main effect function and the interaction function belong to the same family). We also consider two additional scenarios where $k_{main} \neq k_{int}$. 

\vspace{1.6em}
\textbf{The data generation mechanisms are:}
\begin{itemize}
\item Three simple (non)linear kernels that can be sufficiently modeled using finite-dimensional, parametric functions.
\begin{itemize}
\item \code{linear}: $k_{main}$ is a \code{polynomial} kernel with degree 1.
\item \code{quadratic}: $k_{main}$ is a \code{polynomial} kernel with degree 2.
\item \code{cubic}: $k_{main}$ is a \code{polynomial} kernel with degree 3.
\end{itemize}
\item Four flexible nonlinear kernels that each represent the space of all continuous functions with a prespecified set of mathematical properties (e.g., differentiability or complexity). Data generated by these kernels are usually more difficult to model.
\begin{itemize}
\item \code{rbf_1}: A Gaussian RBF kernel with length-scale $1$. This kernel represents the space of functions that are \textit{smooth} (i.e., infinitely differentiable) and have reasonable \textit{complexity} (i.e., does not have fast-varying fluctuations that are difficult to model). 
%Below two $k_{main}$'s extends \code{rbf_1} in terms of %either smoothness or complexity.
\item \code{rbf_0.5}: A Gaussian RBF kernel with length-scale $0.5$. Compared to \code{rbf_1}, \code{rbf_0.5} has the same degree of smoothness but is more complex, i.e., has fast-varying local fluctuations.
\item \code{matern_2.5_1}: A Mat\'{e}rn $\frac{5}{2}$ kernel with length-scale $1$. Compared to \code{rbf_1},\\ \code{matern_2.5_1} has the same degree of complexity but is less smooth, in the sense that it represents the space of twice-differentiable functions,  but is not necessarily infinitely differentiable.
\item \code{matern_1.5_0.5}: A Mat\'{e}rn $\frac{3}{2}$ kernel with length-scale $0.5$. Compared to \code{matern_2.5_1}, \code{matern_1.5_0.5} is more complex but less smooth. It represents the space of once-differentiable functions.
\end{itemize}
\item Two data generation mechanisms where the true main effect $h_1 + h_2$ and the true interaction effect $h_{12}$ are generated from two separate  kernel families (i.e., $k_{main}\neq k_{int}$). The data generating mechanisms are
\begin{itemize}
\item \code{quadratic_rbf}: The main effects are generated from \code{polynomial} with degree 2, while the interaction effect is generated from \code{rbf} with length-scale 1. This combination generates data under quadratic main effects but a flexible nonlinear interaction:
$$y = x_1^2\beta_1 + x_2^2 \beta_2 + \delta * h_{12}(x_1, x_2) + \epsilon.$$

\item \code{rbf_lnr_0.5}: The main effects are generated from \code{rbf} with length-scale 0.5, while the interaction effect is generated from \code{polynomial} with degree 1. This combination generates data under  flexible nonlinear main effects but with a linear interaction:
$$y = h_1(x_1) + h_2(x_2) + \delta * x_1 x_2 + \epsilon.$$
\end{itemize}
\end{itemize}

\textbf{The types of model libraries $k_{model}$ considered are}: 
\begin{itemize}
\item \textbf{Polynomial}: A library of three parametric, \code{polynomial} kernels with degree $p=1, 2, 3$. This library represents a parametric model with polynomial nonlinearity.
\item \textbf{RBF} A library of three nonparametric, \code{rbf} kernels with length-scale $l=0.5, 1, 2$. This library represents a nonparametric model with a high degree of smoothness (i.e., infinitely differentiable) that can incorporate more general types of nonlinearity.
\item \textbf{Polynomial+RBF} A library of three \code{polynomial} kernels with $p=1, 2, 3$ and three \code{rbf} kernels with $l=0.5, 1, 2$. This library represents a semi-parametric model with a mixture of parametric and nonparametric kernels.
%\item 3 \code{matern} kernels ($\nu=1/2, 3/2, 5/2$) with $l=1$ and 3 \code{rbf} kernels ($l=0.6, 1, 2$), representing a very flexible nonparametric model that includes both non-smooth (once- and twice-differentiable) and smooth functions.
\end{itemize}
To understand how the choices of model-selection criteria and ensemble strategy impact the model performance, for each combination of data generation  mechanism and model library, we estimate the null model $y=h_1(x_1) + h_2(x_2)$ under all possible choices of model selection criteria (\code{loocv, AIC, AICc, BIC, GCV, GCVc, gmpml}) and ensemble strategy (\code{avg, exp, stack}). In Sections \ref{sec:simu_kern_lib}-\ref{sec:simu_model_crit}, unless otherwise specified, we report results using \code{loocv} for tuning parameter selection and \code{stack} for the ensemble strategy, which corresponds to the default setting in \pkg{CVEK}. In general, \code{loocv} guarantees correct Type I error, except for Cubic data, which is hard to fit. Also, while correct Type I error is guaranteed, \code{stack} leads to better power. We provide the results for the other settings in Appendix \ref{appendix_A}.
%and two types of interaction kernel (\code{linear, ensemble}) resulting in 18 null model estimates for each of the $18 \times 7 \times 3 = 378$ combinations, and $378 \times 2 = 756$ testing results. 

Figures 1-4 present the results. They show the estimated $\hat{P}(p<0.05)$ (y-axis) as a function of interaction strength $\delta \in [0, 1]$ (x-axis). Each panel in the figure represents the result for a specific data generating mechanism (Linear, Quadratic, Cubic, RBF $l=1$, RBF $l=0.5$, Mat\'{e}rn $\nu=5/2$ with $l=1$ and Mat\'{e}rn $\nu=3/2$ with $l=0.5$), while the different lines represent results from different modeling choices (e.g., choice of null model model library, type of hypothesis test, choice of model selection criteria, etc). 

\textbf{Summary of Recommendations}
Our key recommendations are (1) for kernel library design, it is beneficial to include flexible nonlinear kernels such as the \textbf{RBF} (Section \ref{sec:simu_kern_lib}), (2) for hypothesis testing, use the parametric bootstrap test and the  linear alternative kernel when the sample size is small (Section \ref{sec:simu_test_type}), and (3) choose \code{loocv} as the model selection criterion and \code{stack} as the ensemble strategy (Section \ref{sec:simu_model_crit}).

\subsection{Impact of model library for null model}
\label{sec:simu_kern_lib}

\begin{figure}[ht]
\begin{center}
\includegraphics[width=1\columnwidth]{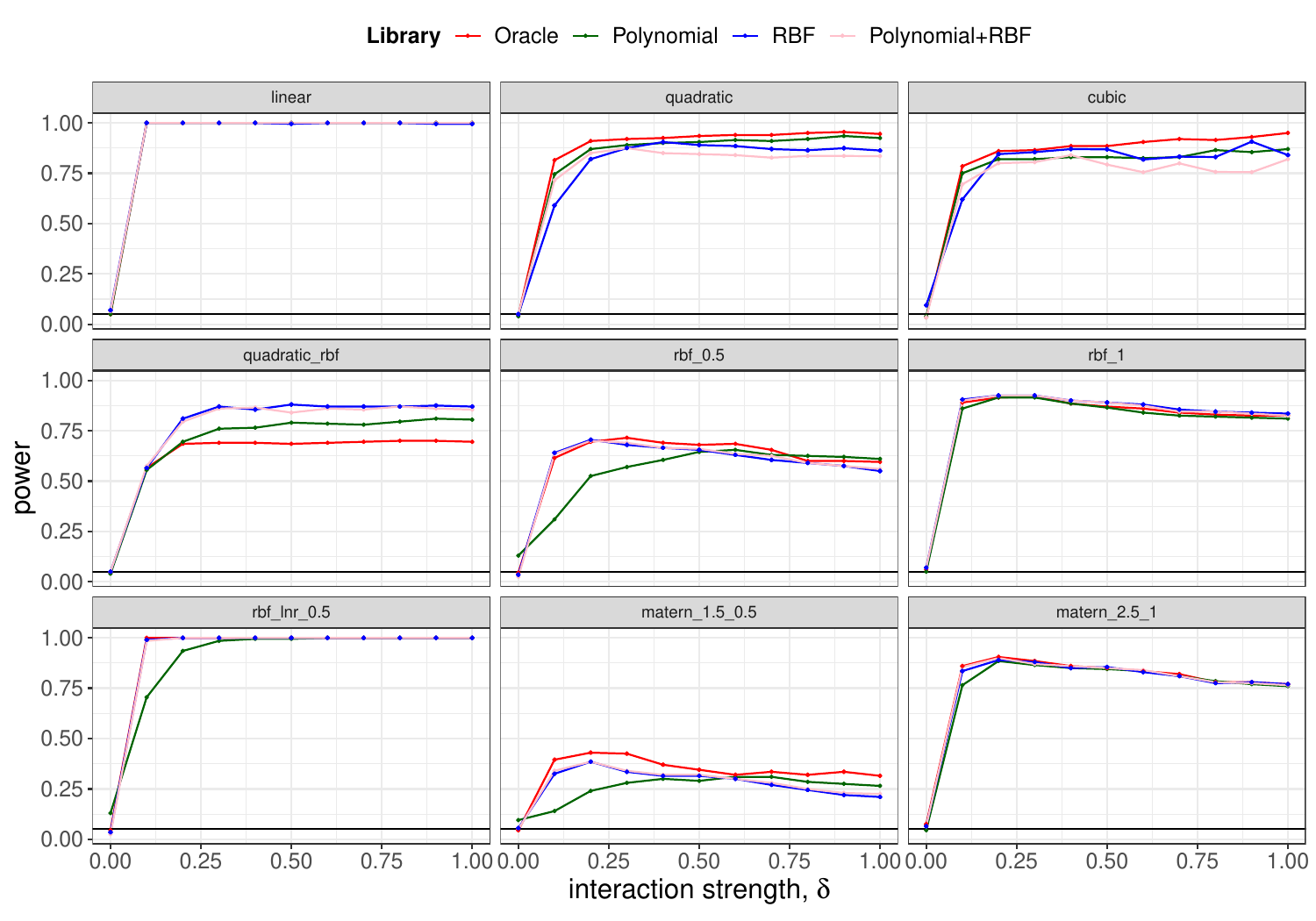} 
\caption{Power of hypothesis test using the true model (Oracle) and the three $k_{model}$'s, fixing the tuning parameter selection method to \code{loocv}, ensemble strategy to \code{stack}, and the test type to bootstrap test with linear alternative kernel.}
\label{fig:1}
\end{center}
\end{figure}

Figure \ref{fig:1} compares the performance of the hypothesis test constructed when the null model is fixed to match the true model (\textbf{Oracle}), or estimated using one of the three different types of model libraries (\textbf{Polynomial}, \textbf{RBF}, \textbf{Polynomial + RBF}). Here, the test is 
based on the bootstrap test with linear alternative kernel. Results for other test types are reported in the Appendix \ref{appendix_A}.

Generally speaking, when all kernels in the model library are parametric (i.e., \textbf{Polynomial} library), the resulting test is more powerful for polynomial data (e.g., \code{linear}, \code{quadratic}, \code{cubic} in Figure \ref{fig:1}), but loses power for data with nonlinearities (e.g., \code{quadratic_rbf}, \code{rbf_0.5}, \code{rbf_lnr_0.5}, \code{matern_1.5_0.5} in Figure \ref{fig:1}). On the other hand, model libraries that involves nonparametric kernels (e.g., \textbf{RBF} library and \textbf{Polynomial + RBF} library) lead to tests that yield slightly less power for polynomial data, but more more powerful for data with more complex nonlinearities (e.g., \code{quadratic_rbf}, \code{rbf_0.5}, \code{rbf_lnr_0.5}, \code{matern_1.5_0.5} in Figure \ref{fig:1}). Comparing the test power between a purely non-parametric model library (\textbf{RBF}) versus a semi-parametric model library (\textbf{Polynomial + RBF}), \textbf{RBF} performs slightly better than \textbf{Polynomial + RBF} library, especially when data comes from quadratic main effect with RBF interaction. This is likely due to the fact that adding less flexible polynomial kernels doesn't result in much of an increase in bias, but introduces more variance due to the need to estimate additional weights. Consequently, we recommend designing the model library to include the nonparametric kernels (e.g., \textbf{RBF} library or \textbf{Polynomial + RBF} library).

\subsection{Impact of test type and choice for the alternative kernel}
\label{sec:simu_test_type}

\begin{figure}
\begin{center}
\includegraphics[width=1\columnwidth]{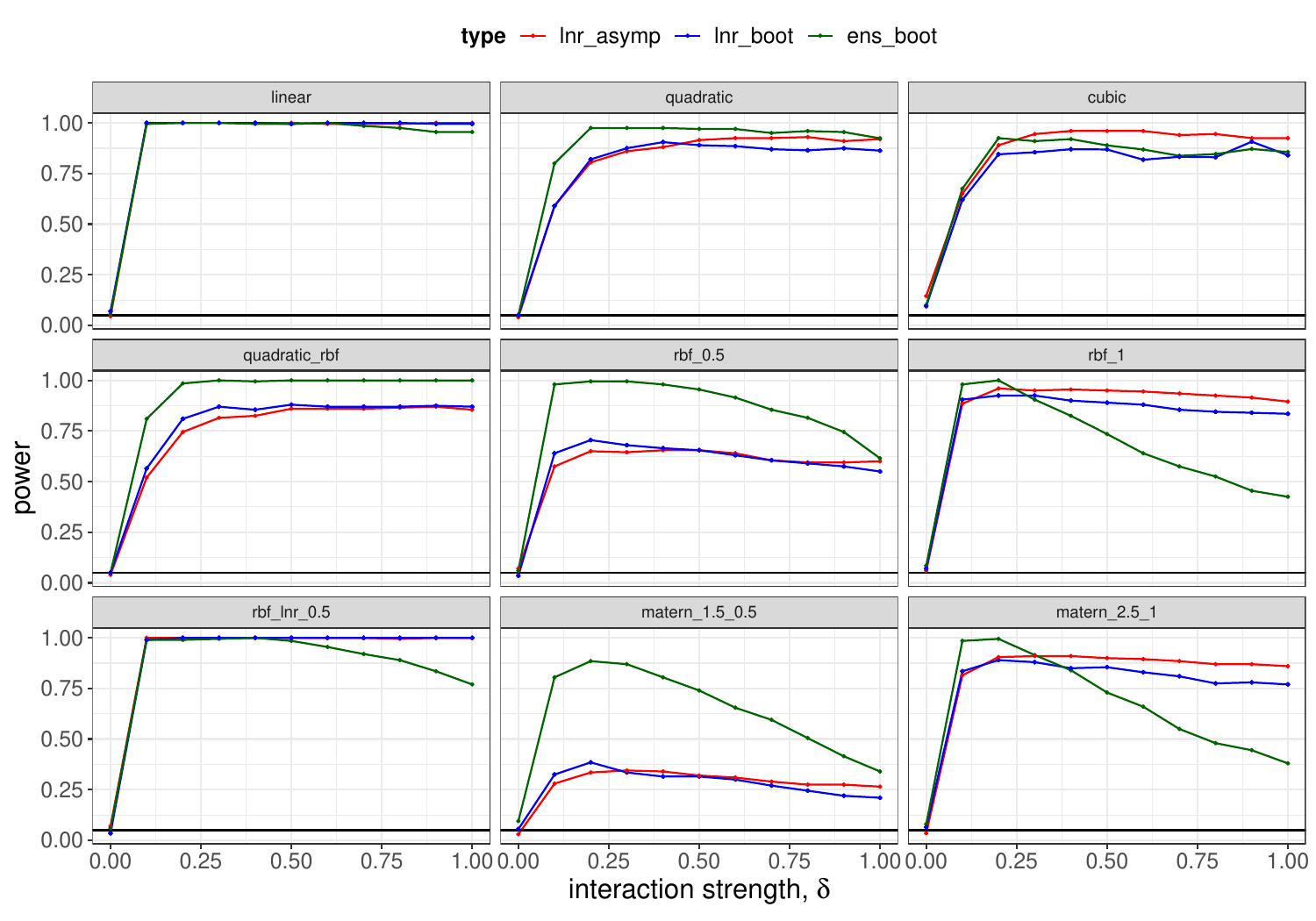} 
\caption{Power of hypothesis test using different types of tests, fixing the tuning parameter selection method to \code{loocv}, ensemble strategy to \code{stack}, and model library for null model to \textbf{RBF}.}
\label{fig:2}
\end{center}
\end{figure}

Figure \ref{fig:2} compares the performance of three types of hypothesis test (\code{lnr_asymp} refers to asymptotic test with linear alternative kernel, \code{lnr_boot} which is the bootstrap test with linear alternative kernel and \code{ens_boot} which is the bootstrap test with ensemble alternative kernel) under different data generation mechanisms. We fix the the model library for null model to \textbf{RBF}.

We first compare the asymptotic to the bootstrap test under the linear alternative kernel. Here, the asymptotic test is observed to be more powerful when data is generated from parametric models with polynomial nonlinearity (e.g., \code{linear}, \code{quadratic}, \code{cubic} in Figure \ref{fig:2}), but has difficulty in  guaranteeing correct Type I error under more complex data (e.g., the \code{cubic}). Meanwhile, when data are generated from more flexible kernels (e.g., \code{rbf_0.5}, \code{matern_1.5_0.5} in Figure \ref{fig:2}), the bootstrap test is observed to be slightly more powerful than the asymptotic test. So we recommend choosing the test type to be bootstrap.

Now, fixing the test type to be bootstrap test, we consider the impact of the choice of the alternative kernel on test performance. Specifically, we are interested in whether the test with an adaptively estimated alternative kernel (using the strategy outlined in Section \ref{sec:alt_kernel}) leads to better power compared to the one with linear kernel as the alternative kernel. In general, the test based on the ensemble alternative kernel is more powerful than that generated by the linear alternative kernel, especially when data are generated having high complexity with quickly-varying local fluctuations (e.g., \code{rbf_0.5}, \code{matern_1.5_0.5} in Figure \ref{fig:2}). Interestingly, when the data are simulated under nonlinearities generated from the RBF or the Mat\'{e}rn kernel, the test power sometimes decreases as the data generation mechanism moves away from the null (e.g., \code{rbf_1}, \code{matern_2.5_1} in Figure \ref{fig:2}). The simulation in Appendix \ref{appendix_A} also shows this phenomenon. This result suggests the test statistic \eqref{test_stat} calculated with the ensemble alternative kernel can be unstable and yields an underestimate of the true alternative kernel under the alternative. Since the linear alternative kernel is more stable, we recommend choosing a linear alternative kernel in combination with the parametric bootstrap for testing.

\subsection{Impact of model selection criterion and ensemble strategy}
\label{sec:simu_model_crit}

\begin{figure}
\begin{center}
\includegraphics[width=1\columnwidth]{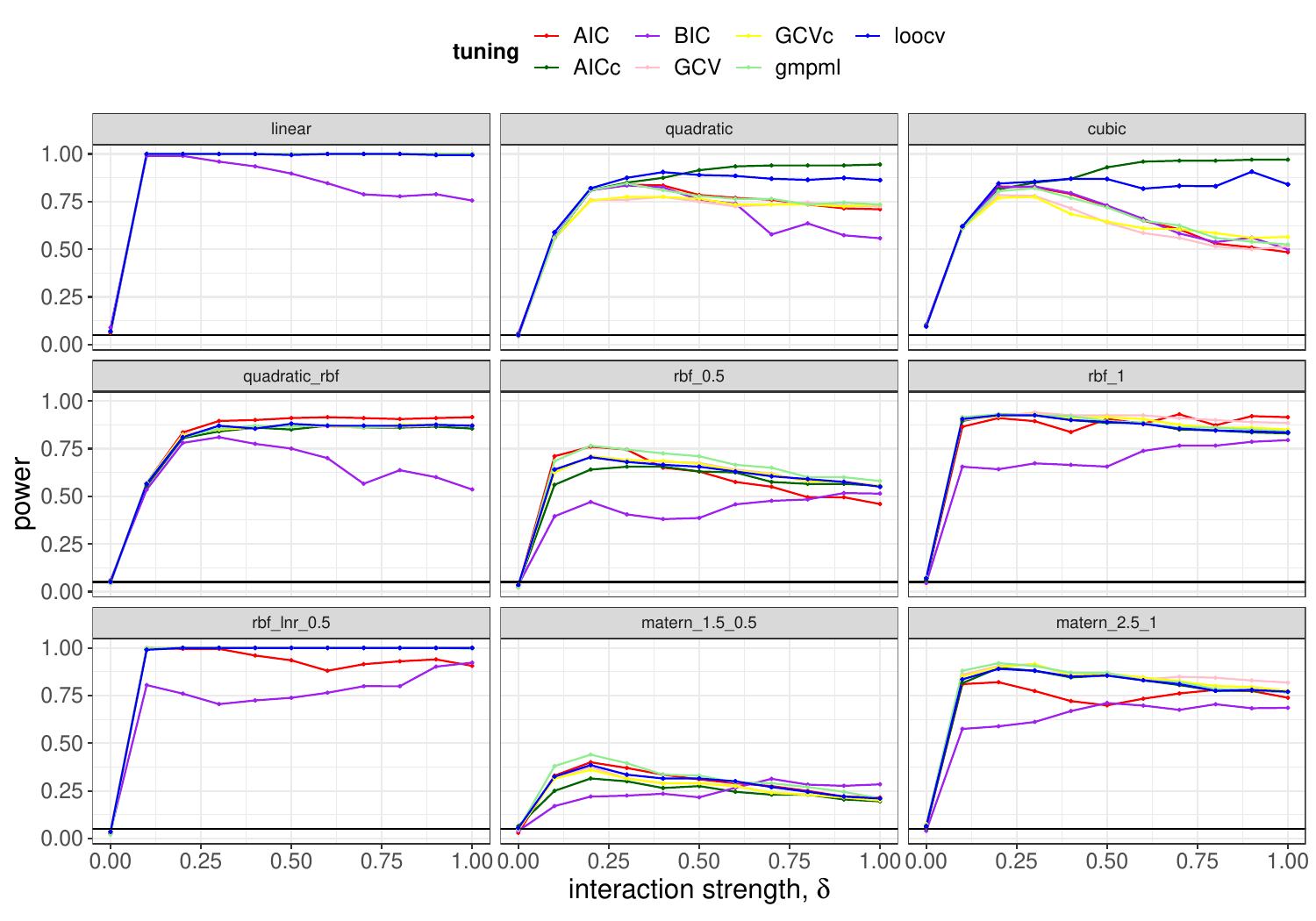} 
\caption{Performance of hypothesis test when null models are selected from different tuning parameter selections, fixing the ensemble strategy to \code{stack}, test type to bootstrap test with linear alternative kernel, and model library for null model to \textbf{RBF}.}
\label{fig:3}
\end{center}
\end{figure}

Figure \ref{fig:3} compares the performance of the proposed hypothesis test when null models are selected from different tuning parameter criteria. This figure presents results when fixing the model library for null model to \textbf{RBF} and test type to bootstrap test with linear alternative kernel. 

Among the seven tuning parameter selection methods (\code{loocv, AIC, AICc, BIC, GCV, GCVc, gmpml}), we notice that \code{loocv} is generally better at guaranteeing correct Type I error, except for Cubic data, which is hard to fit for all selection methods. Furthermore, it is important to note that some conventional hyper-parameter selection criteria (e.g., \code{AIC} in red and \code{BIC} in purple) produces suboptimal tests with inflated Type I error and weak power (e.g., \code{cubic}, \code{quadratic_rbf}, \code{rbf_0.5} in Figure \ref{fig:3}). The issue with test power is especially severe when data comes from polynomial main effect, as the test power is weak and sometimes even decreases as the data generation mechanism moves away from the null (e.g., \code{cubic}, \code{quadratic_rbf} in Figure \ref{fig:3}). This result suggests poor fits of the null model as estimated using AIC/BIC. Consequently, we recommend using \code{loocv} and avoid using AIC/BIC-type criteria under these scenarios.

\begin{figure}
\begin{center}
\includegraphics[width=1\columnwidth]{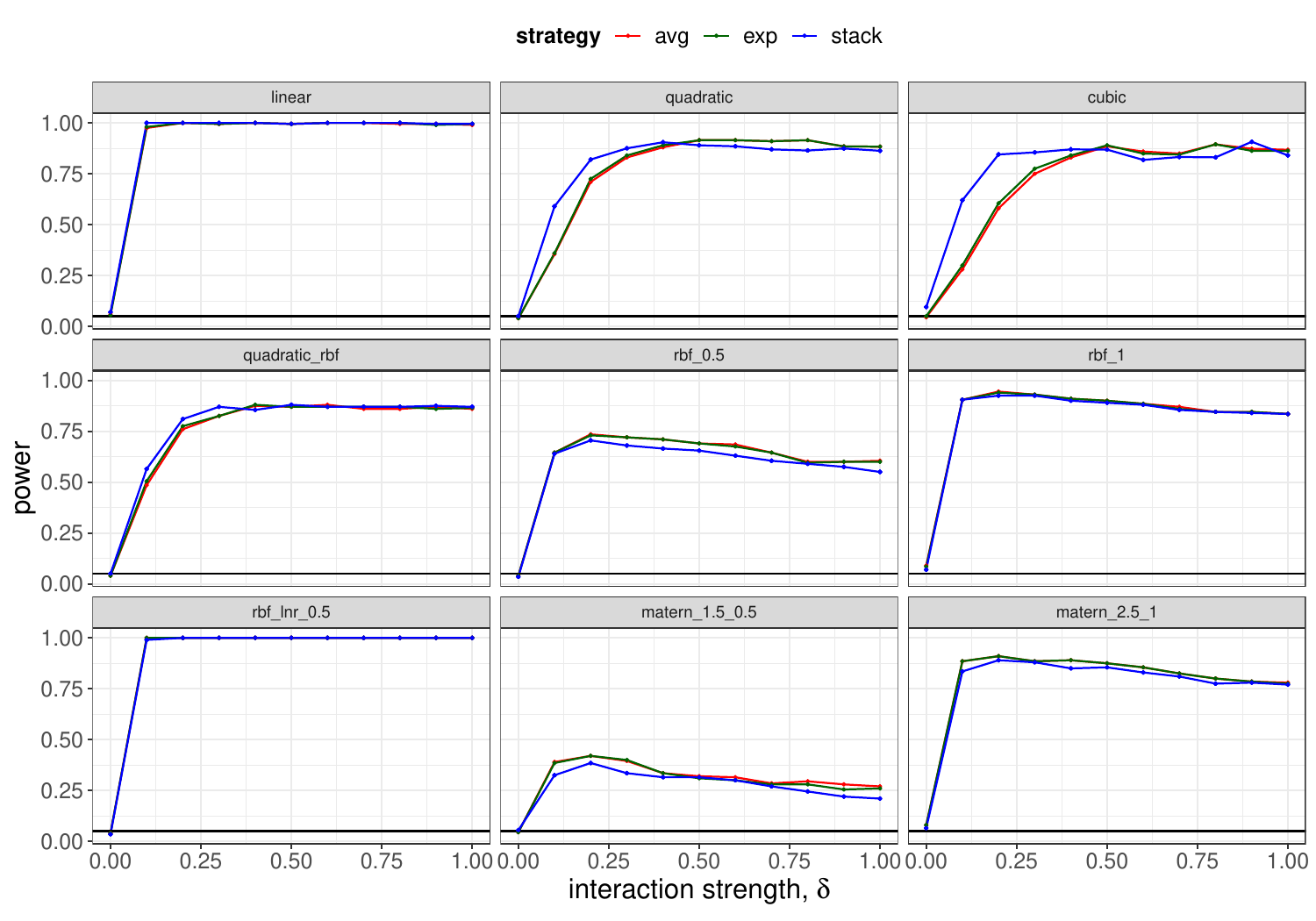} 
\caption{Performance of hypothesis test when null models are selected from different ensemble strategies, fixing the tuning parameter selection method to \code{loocv}, test type to bootstrap test with linear alternative kernel, and model library for null model to \textbf{RBF}.}
\label{fig:4}
\end{center}
\end{figure}

Figure \ref{fig:4} compares the performance of hypothesis test when null models are selected from different ensemble strategies, fixing the model library for null model to \textbf{RBF} and test type to bootstrap test with linear alternative kernel. Interestingly, while Cubic data is hard to fit and \code{stack} is more powerful, \code{avg} and \code{exp} can guarantee correct Type I error. This is likely due to the fact that \code{avg} and \code{exp} have closed-form solutions for weights while the weights of \code{stack} need to be estimated, thus introducing more variance. On the other hand, among the three ensemble strategies, we notice that \code{stack} is generally more powerful than \code{avg} and \code{exp}. Therefore, we recommend choosing \code{stack} as the ensemble strategy.

\section{Conclusion}
\label{sec:conclusion}

In this paper we describe \pkg{CVEK}, a powerful and flexible toolkit in R for learning nonlinear, multivariate feature  effects in limited samples. Given data, \pkg{CVEK} efficiently learns the complex, nonlinear data generation mechanism using an ensemble of kernel machine regressions. The package offers flexible choices for both kernel specification and hyper-parameter tuning for constructing the base models in a given library, as well as different ensemble strategies for constructing the final ensemble. \pkg{CVEK} also offers a suite of hypothesis tests for both the main effects and the interaction effects for the kernel features, and provides an asymptotic approximation and the parametric bootstrap as options  for estimating the null distribution of the resulting test statistic. Further, one can use either the linear alternative kernel or an ensemble alternative kernel to test interaction effect. Through comprehensive simulation, we show that the hypothesis tests offered by \pkg{CVEK} are valid and powerful even under  complex, non-smooth data generating mechanisms where the classical approaches fail. In practice, for a robust option for general purpose use, we recommend users employ an RBF model library for the ensemble, select the hyper-parameter using loocv, and construct the ensemble using cross-validated stacking. For testing, we recommend one chooses the linear alternative kernel in combination with the parametric bootstrap. We encourage practitioners to conduct sensitivity analysis in order to examine the robustness of their conclusions, utilizing the wide range of options provided by \pkg{CVEK}.

%% -- Optional special unnumbered sections -------------------------------------
\newpage
\section*{Tables and Figures} 
\label{sec:append}

\newpage

\begin{algorithm}
\caption{Cross validated ensemble of kernels (CVEK)} 
\begin{algorithmic}[1]
\Procedure{CVEK}{}\newline
\textbf{Input:} A library of kernels $\{k_j\}_{j=1}^K$, Data $(\by, \bx)$\newline
\textbf{Output:} Ensemble Kernel Matrix $\widehat{\bK}$\newline
$\#$ \texttt{Stage 1: Estimate $\lambda$ and CV error for each kernel}
\For{$j = 1$ to $K$}
\State{$\bK_j = \bK_j/\tr(\bK_j)$}
\State{$\widehat{\lambda}_j = \mbox{\textit{argmin}} \; 
\texttt{LOOCV}\Big(\lambda \mid \bK_j \Big)$} 
\State{$\widehat{\xi}_j = 
\texttt{CV}\Big(\widehat{\lambda}_j \mid \bK_j \Big)$}
\EndFor 
\newline
$\#$ \texttt{Stage 2: Estimate ensemble weights $\bu_{K \times 1} = \{u_1, \dots, u_K\}$}
\State{
$\widehat{\bu} = \underset{\bu \in \Delta}{argmin} \; \parallel \sum_{j=1}^K u_j \widehat{\xi}_j\parallel^2 \qquad 
\mbox{where} \quad
\Delta = \{\bu\mid \bu \geq 0, \parallel \bu \parallel_1 = 1 \}$}
\newline
$\#$ \texttt{Stage 3: Assemble the ensemble kernel matrix $\widehat{\bK}_{ens}$}
\State{$\widehat{\bA} = \sum_{j=1}^K 
\widehat{\mu}_j\bA_{\widehat{\lambda}_j, k_j}$}
\State{$\bU_A, \bdelta_{A} = 
\texttt{spectral\_decomp}(\widehat{\bA})$}
\State{$\lambda_{\bK} = min \Big(1,  
(\sum_{k=1}^n \frac{\delta_{A, k}}{1 - \delta_{A,k}})^{-1}, 
min \big(\{\widehat{\lambda}_j\}_{j=1}^K \big)
\Big)$}
\State{$\widehat{\bK} = 
\lambda_{\bK} * 
\widehat{\bU}_A \; 
diag\Big( \frac{\delta_{A,k}}{1 - \delta_{A,k}} \Big) \;
\widehat{\bU}_A^\top $}
\EndProcedure
\end{algorithmic}
\end{algorithm}

\begin{algorithm}
\caption{Variance component test for $h \in \Hsc_0$} 
\begin{algorithmic}[1]
\Procedure{VCT FOR INTERACTION}{}\newline
\textbf{Input:} Null Kernel Matrix $\bK_0$, Derivative Kernel Matrix $\partial \bK_0$, Data $(y, \bx)$\newline
\textbf{Output:} Hypothesis Test p-value $p$\newline
$\#$ \texttt{Stage 1: Estimate Null Model using REML}
\State{$(\hat{\bmu}, \hat{\tau}, \hat{\sigma}^2)=argmax L_{REML}(\bmu, \tau, \sigma^2\mid\bK_0)$}
\newline
$\#$ \texttt{Stage 2: Compute Test Statistic and Null Distribution Parameters}
\State{$\hat{T}_0=\hat{\tau}*(\by-\bX\hat{\bbeta})^\top \bV_0^{-1}\partial \bK_0\bV_0^{-1}(\by-\bX\hat{\bbeta})$}
\State{$\hat{\kappa}=\hat{\bI}_{\delta\delta}/[\hat{\tau}*tr(\bV_0^{-1}\partial \bK_0)]$, \quad $\hat{\nu}=[\hat{\tau}*tr(\bV_0^{-1}\partial \bK_0)]^2/(2*\hat{\bI}_{\delta\theta})$}
\newline
$\#$ \texttt{Stage 3: Compute p-value and reach conclusion}
\State{$p=P(\hat{\kappa} \chi_{\hat{\nu}}^2>\hat{T})=P(\chi_{\hat{\nu}}^2>\hat{T}/\hat{\kappa})$}
\EndProcedure
\end{algorithmic}
\end{algorithm}

\begin{algorithm}
\caption{Parametric bootstrap test} 
\begin{algorithmic}[1]
\Procedure{Parametric Bootstrap Test}{}\newline
\textbf{Input:} Null Kernel Matrix $\bK_0$, Derivative Kernel Matrix $\partial \bK_0$, Data $(\by, \bx)$\newline
\textbf{Output:} Hypothesis Test p-value $p$\newline
$\#$ \texttt{Stage 1: Estimate Null Model using Gaussian Process Regression}
\State{$\hat{\bmu}=\bA_0\by, \quad \hat{\sigma}^2=\frac{\by^\top (\bI-\bA_0)\by}{n-tr(\bA_0)}, \quad \hat{\tau}$}
\newline
$\#$ \texttt{Stage 2: Sample response from the fitted model obtain in Step 1}
\newline
$\#$ \texttt{and compute the test statistic based on fitting the alternative}
\newline
$\#$ \texttt{model, repeat for $B$ times}
\For{$b= 1$ to $B$}
\State{$\by^\star=\hat{\bmu}+\bepsilon,\quad \bepsilon \sim \mathrm{N}(0, \hat{\sigma}^2)$}
\State{$\hat{T}_{0b}=\hat{\tau}*(\by^\star-\hat{\bmu})^\top \bV_0^{-1}\partial \bK_0\bV_0^{-1}(\by^\star-\hat{\bmu})$} 
\EndFor 
\newline
$\#$ \texttt{Stage 3: Compute the test statistic for the original data, based}
\newline
$\#$ \texttt{on fitting the alternative hypothesis model}
\State{$\hat{T}_0=\hat{\tau}*(\by-\hat{\bmu})^\top \bV_0^{-1}\partial \bK_0\bV_0^{-1}(\by-\hat{\bmu})$}
\newline
$\#$ \texttt{Stage 4: Compute p-value and reach conclusion}
\State{$p=\frac{1}{B}\sum_{b=1}^BI(\hat{T}_{0b}>\hat{T}_0)$}
\EndProcedure
\end{algorithmic}
\end{algorithm}

\clearpage
\section*{Computational details}
The results in this paper were obtained using R 3.6.1 with the \pkg{CVEK} 0.1-2 package.

%% -- Bibliography -------------------------------------------------------------
%% - References need to be provided in a .bib BibTeX database.
%% - All references should be made with \cite, \citet, \citep, \citealp etc.
%%   (and never hard-coded). See the FAQ for details.
%% - JSS-specific markup (\proglang, \pkg, \code) should be used in the .bib.
%% - Titles in the .bib should be in title case.
%% - DOIs should be included where available.

\bibliography{refs}

%% -- Appendix (if any) --------------------------------------------------------
%% - After the bibliography with page break.
%% - With proper section titles and _not_ just "Appendix".

\newpage
\begin{appendix}

\newpage
\section{Further simulation results under the settings in Section 5} 
\renewcommand\thefigure{A.\arabic{figure}}    
\setcounter{figure}{0} 
\label{appendix_A}
\subsection{Performances of different libraries combined with different testing types} 
\begin{figure}[ht]
\begin{center}
\includegraphics[width=1\columnwidth]{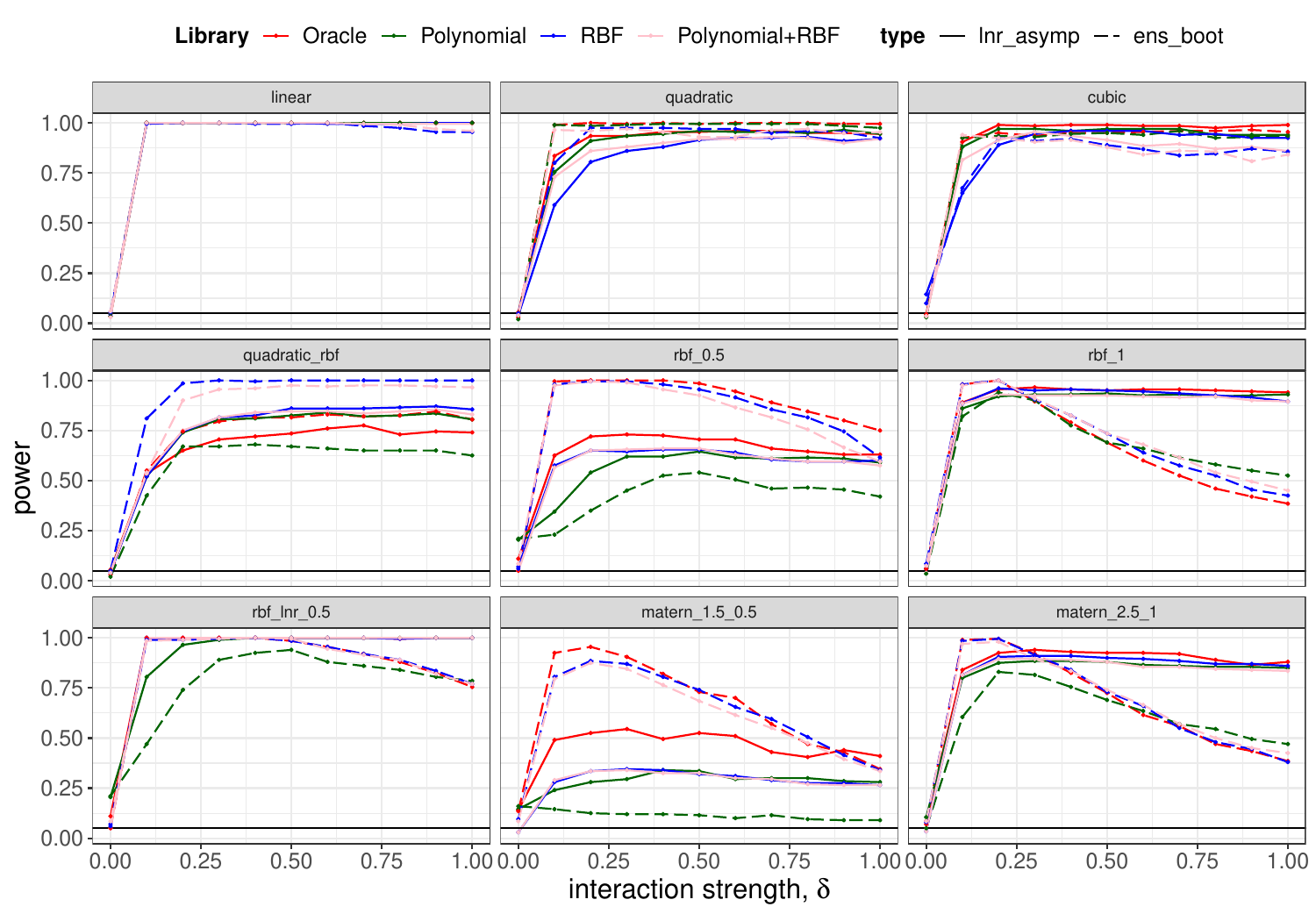} 
\caption{Power of hypothesis test using different libraries combined with different types of tests, fixing the tuning parameter selection method to \code{loocv} and ensemble strategy to \code{stack}.}
\end{center}
\end{figure}

\newpage
\subsection{Performances of different libraries combined with different ensemble strategies} 
\begin{figure}[ht]
\begin{center}
\includegraphics[width=1\columnwidth]{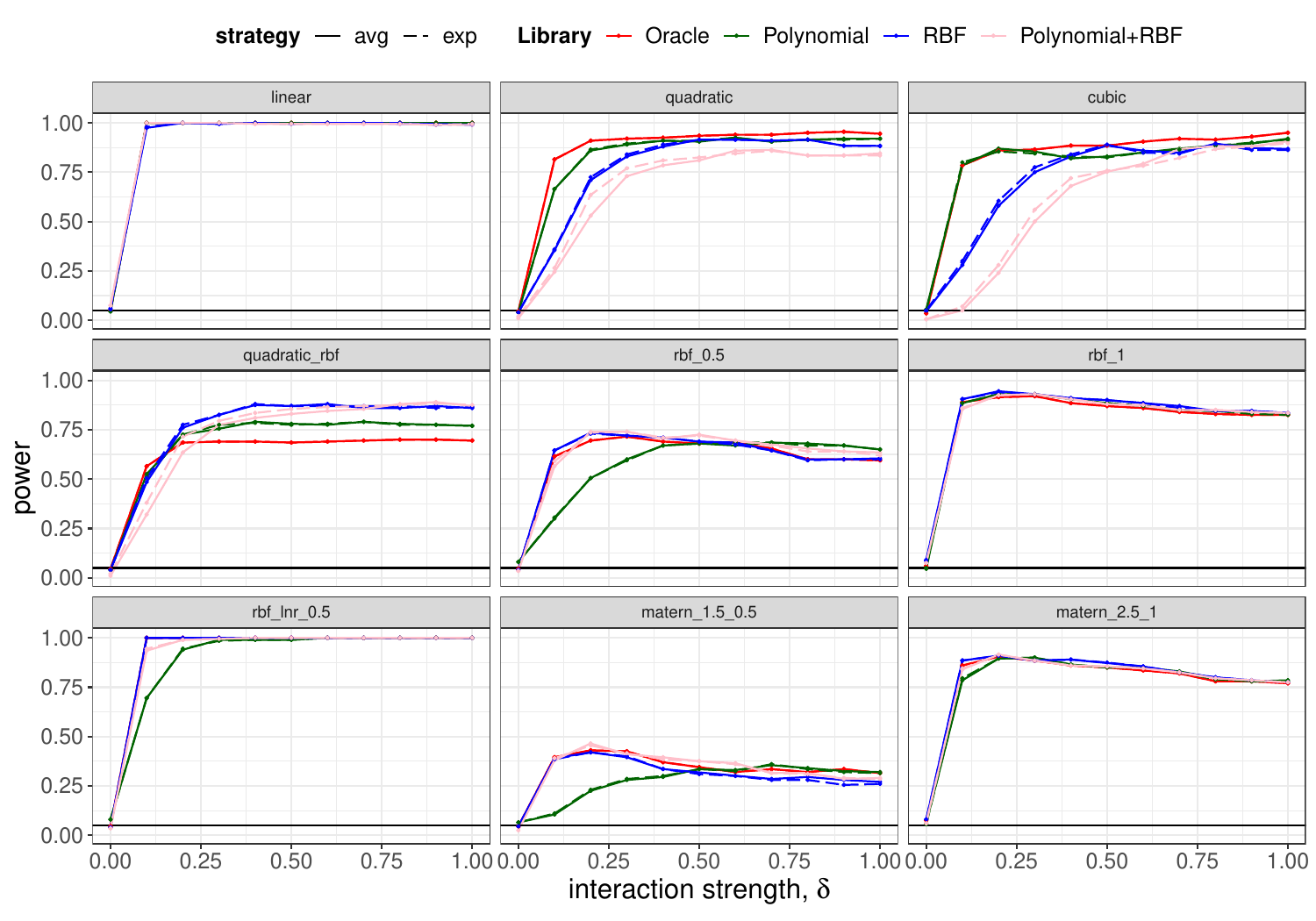} 
\caption{Power of hypothesis test using different libraries combined with different ensemble strategies, fixing the test type to bootstrap test with linear alternative kernel and tuning parameter selection method to \code{loocv}.}
\end{center}
\end{figure}

\newpage
\subsection{Performances of different testing types combined with different ensemble strategies} 
\begin{figure}[ht]
\begin{center}
\includegraphics[width=1\columnwidth]{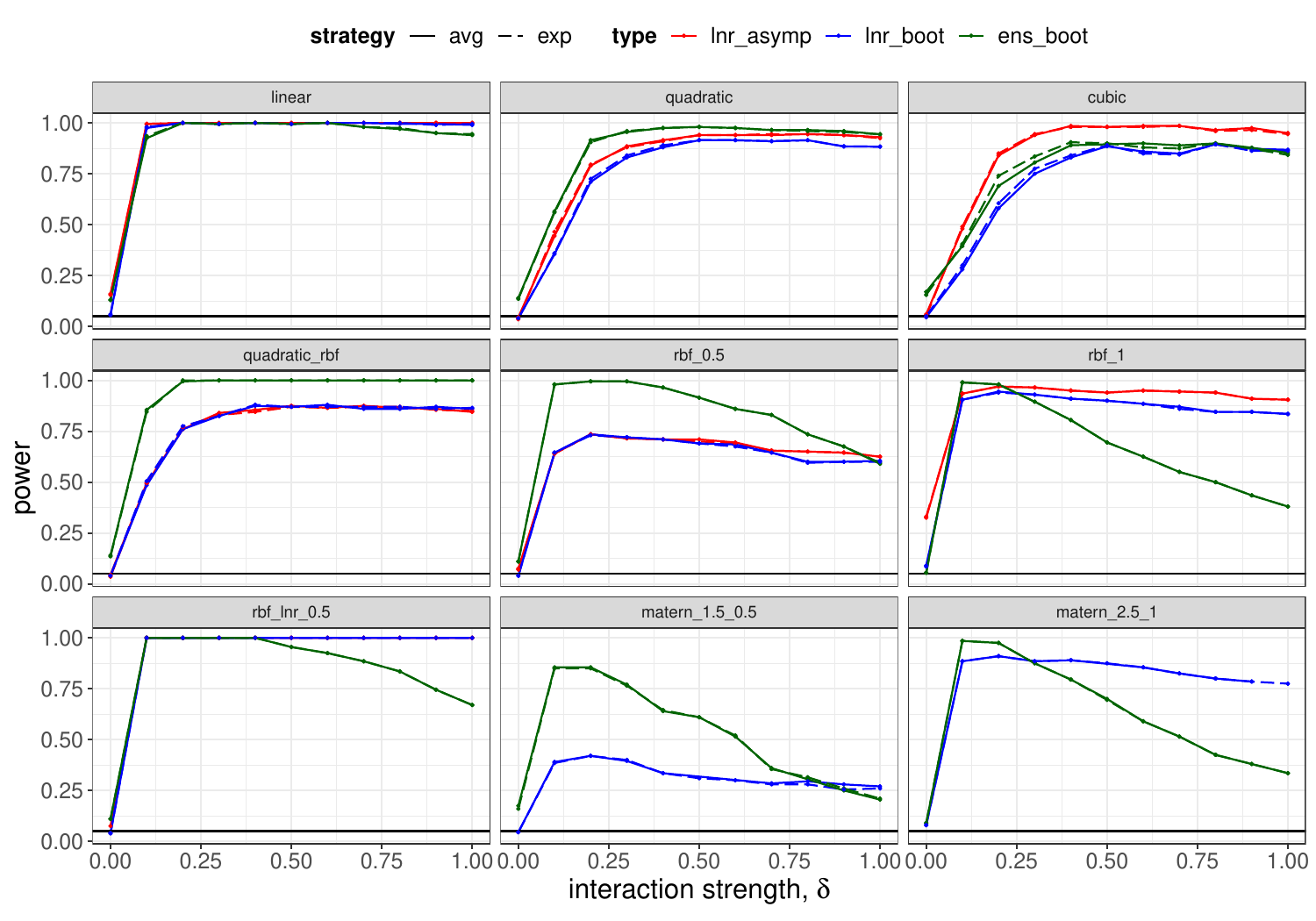} 
\caption{Power of hypothesis test using different testing types combined with different ensemble strategies, fixing the model library for null model to \textbf{RBF} and tuning parameter selection method to \code{loocv}.}
\end{center}
\end{figure}

\newpage
\subsection{Performances of different tuning parameter selections combined with different testing types} 
\begin{figure}[ht]
\begin{center}
\includegraphics[width=1\columnwidth]{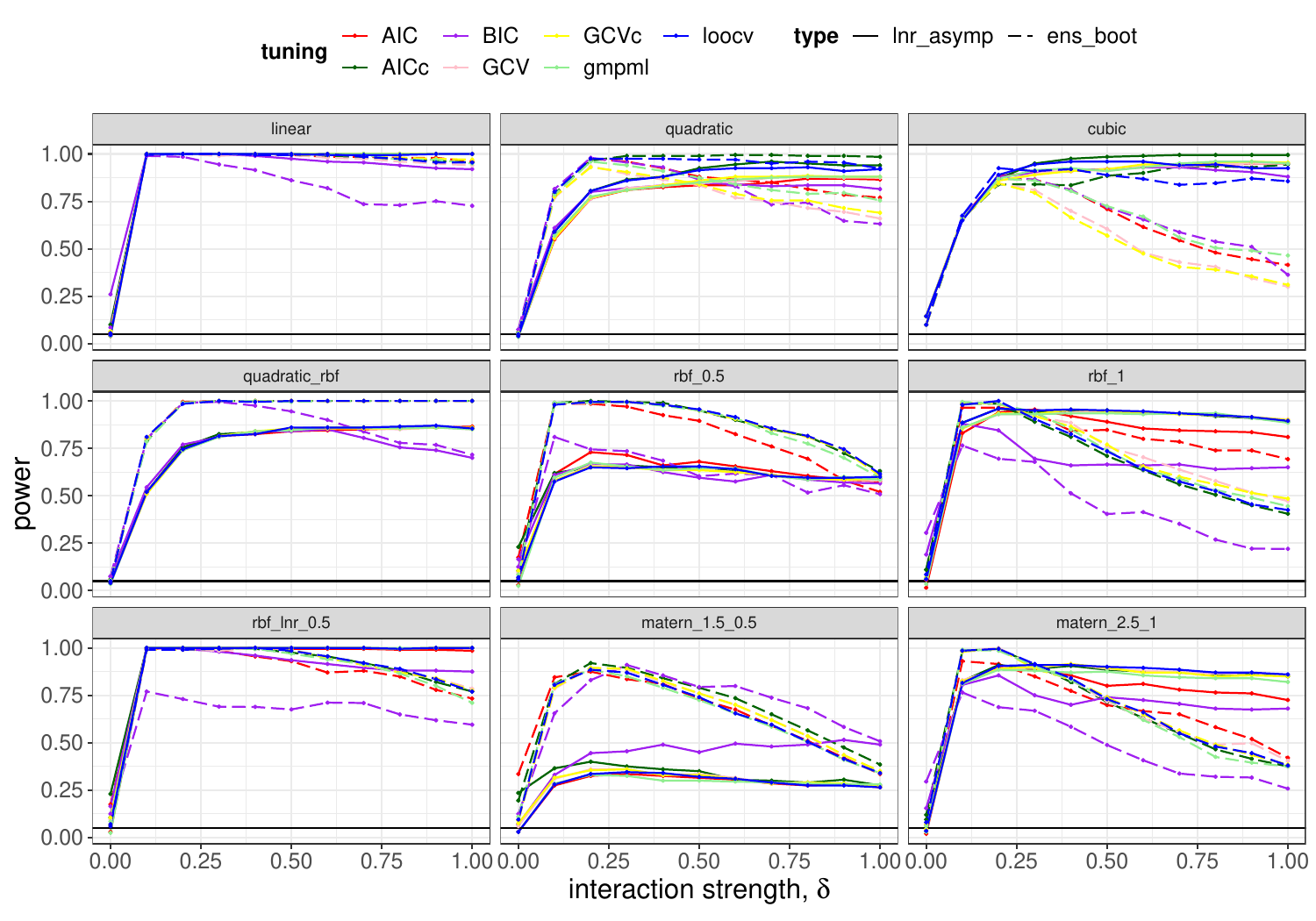} 
\caption{Power of hypothesis test using different tuning parameter selections combined with different types of tests, fixing the model library for null model to \textbf{RBF} and ensemble strategy to \code{stack}.}
\end{center}
\end{figure}

\newpage
\subsection{Performances of different beta's of exponential weighting} 
Below shows the performances under three different beta's of exponential weighting:\\ $\text{min}\{RSS\} _{d=1}^D/10$, $\text{median}\{RSS\} _{d=1}^D$ and $\text{max}\{RSS\} _{d=1}^D * 2$. Here $\{RSS\} _{d=1}^D$ are the set of residual sum of squares of $D$ base kernels. We can see that their performances are quite similar.
\begin{figure}[ht]
\begin{center}
\includegraphics[width=1\columnwidth]{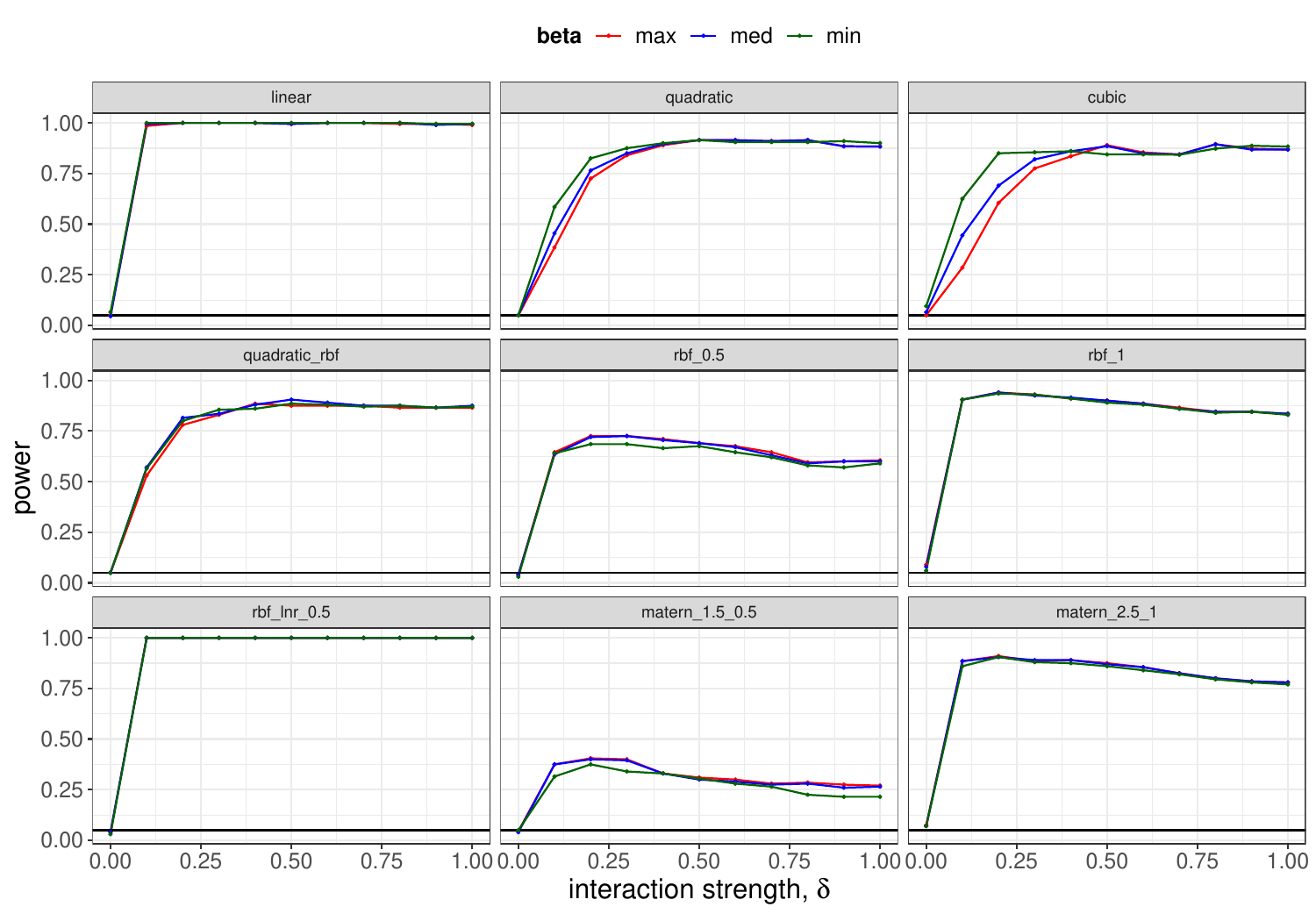} 
\caption{Power of hypothesis test using different beta's of exponential weighting, fixing the model library for null model to \textbf{RBF}, test type to bootstrap test with linear alternative kernel, tuning parameter selection method to \code{loocv} and ensemble strategy to \code{stack}.}
\end{center}
\end{figure}

% \newpage
% \section{Higher-Dimensional and correlated settings} 
% \renewcommand\thefigure{B.\arabic{figure}}    
% \setcounter{figure}{0} 
% \label{appendix_B}
% \subsection{Performances of different libraries under higher-dimensional setting} 
% \begin{figure}[ht]
% \begin{center}
% \includegraphics[width=1\columnwidth]{./plots/d6_library} 
% \caption{Power of hypothesis test using the true model (Oracle) and the three $k_{model}$'s when $p_1=p_2=6$ and $n=500$, fixing the tuning parameter selection method to \code{loocv}, ensemble strategy to \code{stack}, and the test type to bootstrap test with linear alternative kernel.}
% \end{center}
% \end{figure}

\newpage
\section{Correlated settings} 
\renewcommand\thefigure{B.\arabic{figure}}    
\setcounter{figure}{0} 
\label{appendix_B}
\subsection{Performances of different libraries under correlated setting (without between group correlation)}
\begin{figure}[ht]
\begin{center}
\includegraphics[width=1\columnwidth]{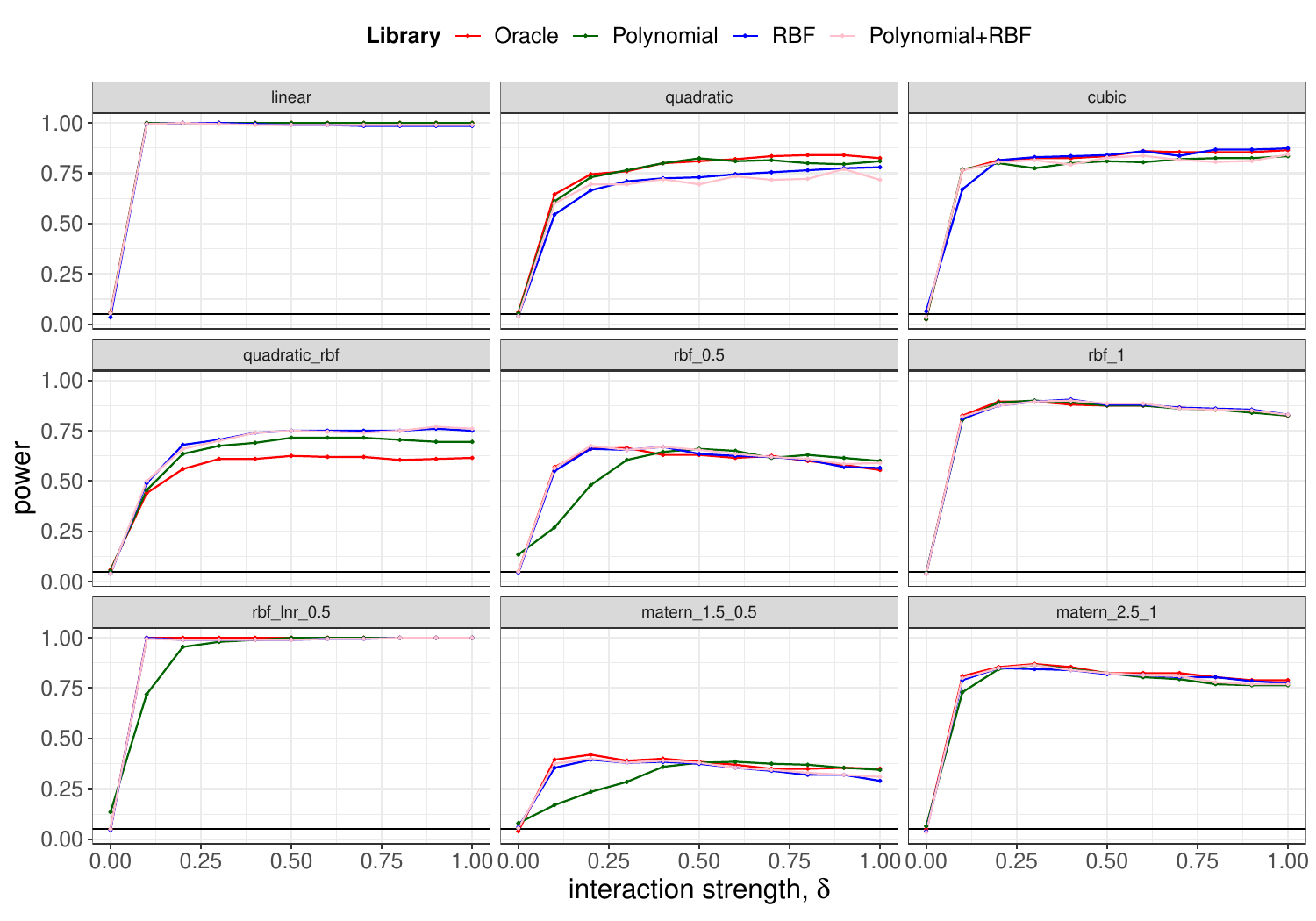} 
\caption{Power of hypothesis test using the true model (Oracle) and the three $k_{model}$'s when within group correlation coefficients are 0.3 and 0.7 respectively, and between group correlation coefficient is 0, fixing the tuning parameter selection method to \code{loocv}, ensemble strategy to \code{stack}, and the test type to bootstrap test with linear alternative kernel.}
\end{center}
\end{figure}

\newpage
\subsection{Performances of different testing types under correlated setting (without between group correlation)} 
\begin{figure}[ht]
\begin{center}
\includegraphics[width=1\columnwidth]{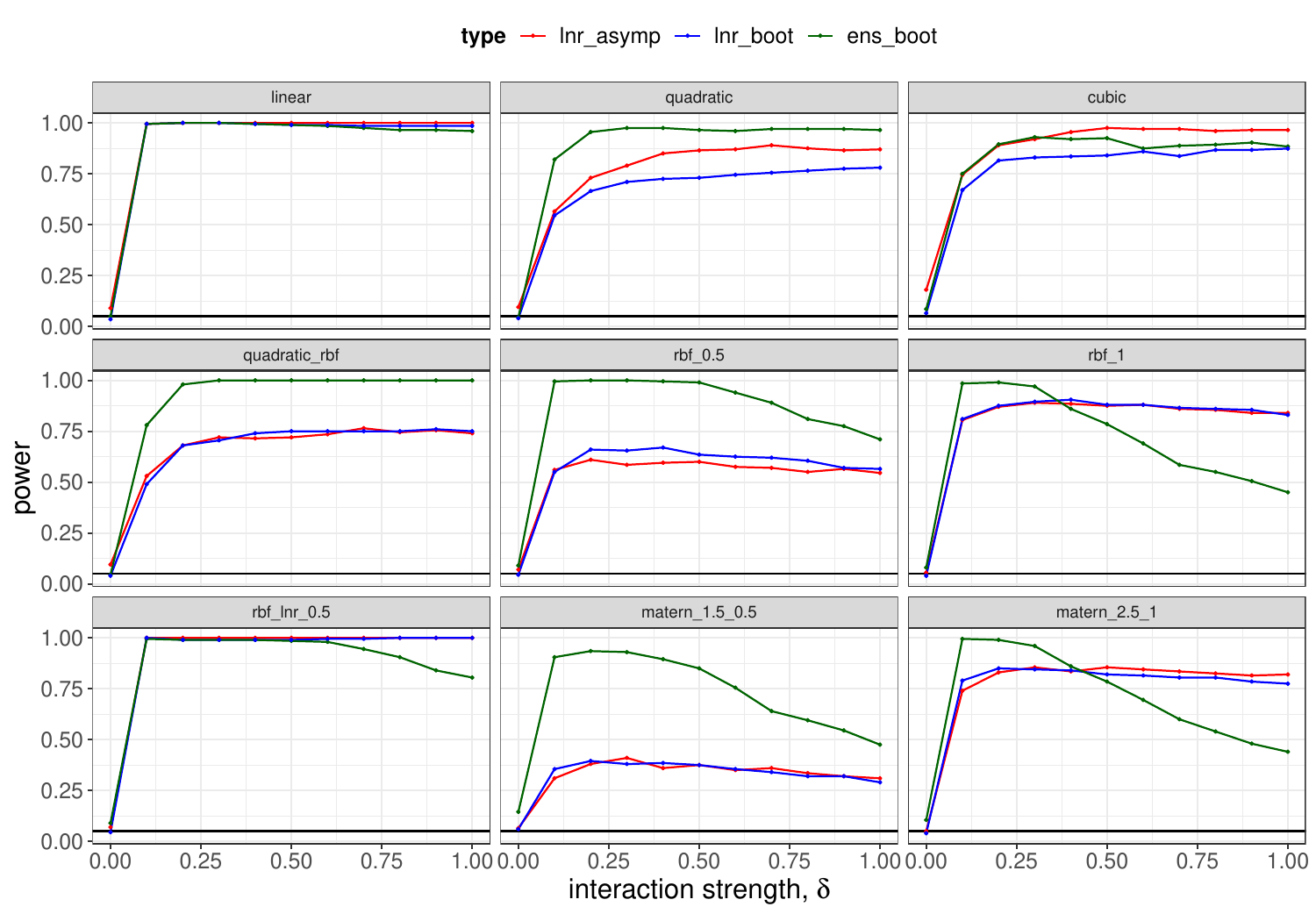} 
\caption{Power of hypothesis test using different types of tests when within group correlation coefficients are 0.3 and 0.7 respectively, and between group correlation coefficient is 0, fixing the tuning parameter selection method to \code{loocv}, ensemble strategy to \code{stack} and model library for null model to \textbf{RBF}.}
\end{center}
\end{figure}

\newpage
\subsection{Performances of different libraries under correlated setting (with between group correlation)}
\begin{figure}[ht]
\begin{center}
\includegraphics[width=1\columnwidth]{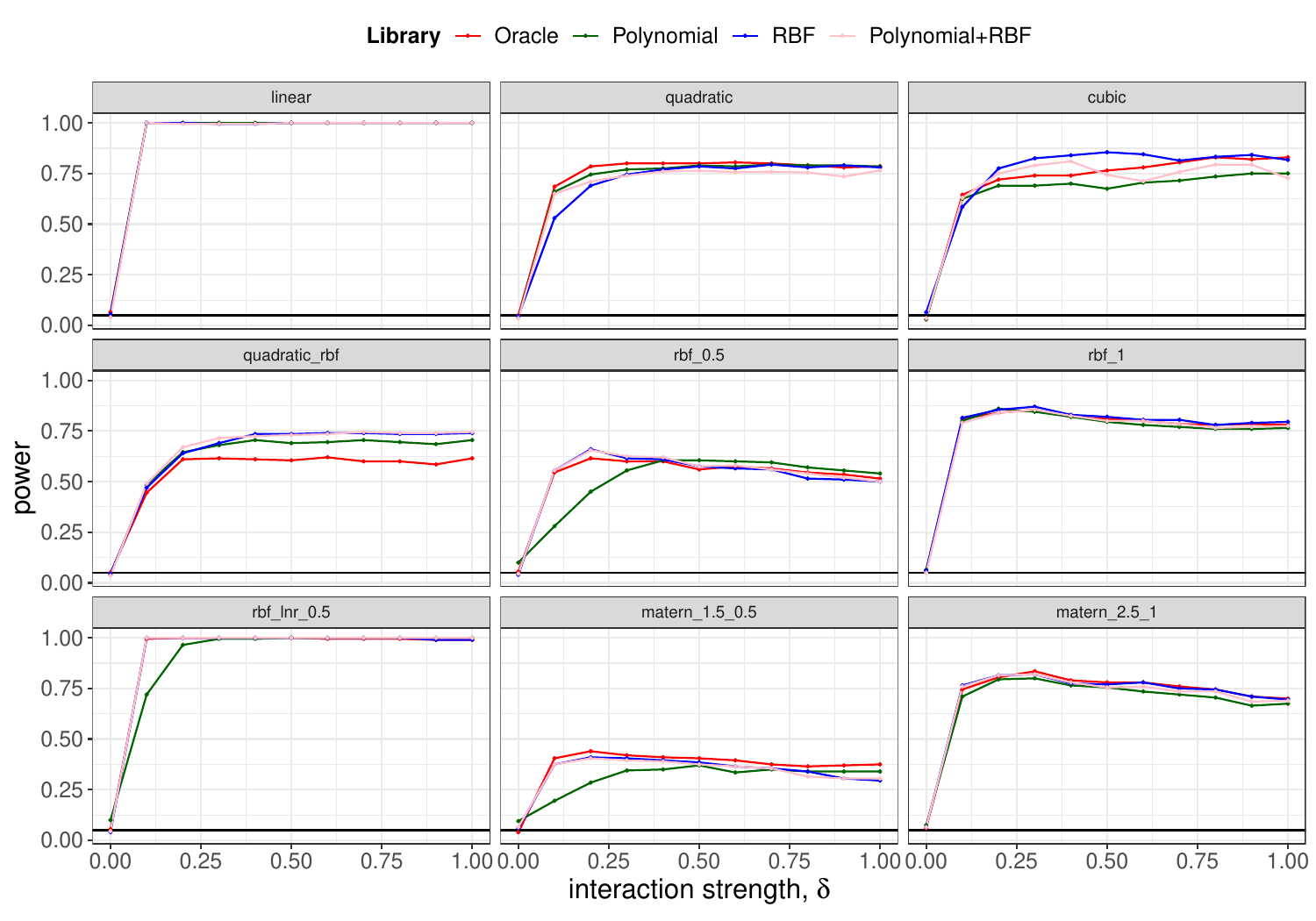} 
\caption{Power of hypothesis test using the true model (Oracle) and the three $k_{model}$'s when within group correlation coefficients are 0.3 and 0.7 respectively, and between group correlation coefficient is 0.2, fixing the tuning parameter selection method to \code{loocv}, ensemble strategy to \code{stack} and the test type to bootstrap test with linear alternative kernel.}
\end{center}
\end{figure}

\newpage
\subsection{Performances of different testing types under correlated setting (with between group correlation)} 
\begin{figure}[ht]
\begin{center}
\includegraphics[width=1\columnwidth]{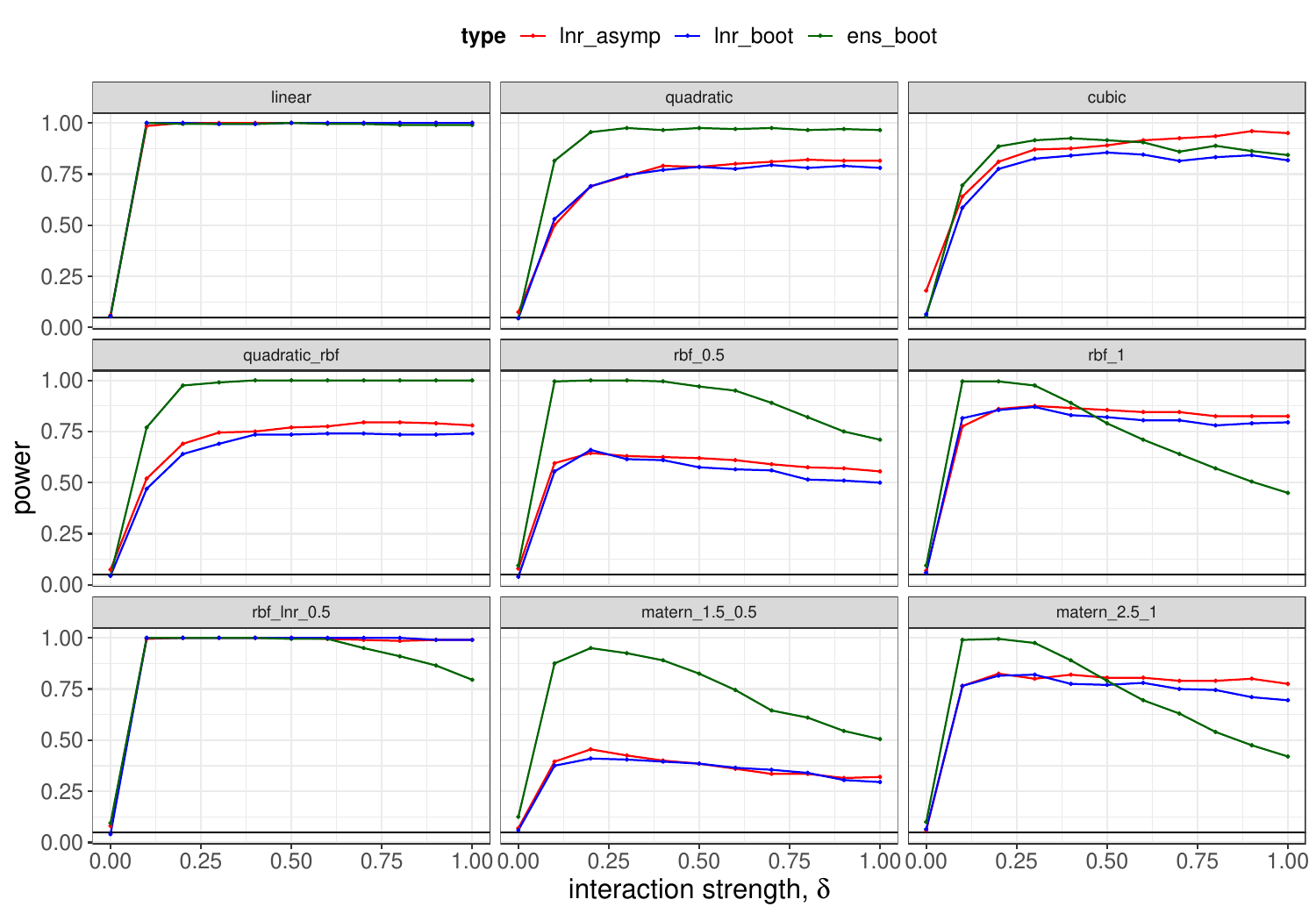} 
\caption{Power of hypothesis test using different types of tests when within group correlation coefficients are 0.3 and 0.7 respectively, and between group correlation coefficient is 0.2, fixing the tuning parameter selection method to \code{loocv}, ensemble strategy to \code{stack} and model library for null model to \textbf{RBF}.}
\end{center}
\end{figure}

\newpage
\section{Backfitting algorithm for multiple kernels}
\begin{algorithm}
\caption{Backfitting algorithm for multiple kernels} 
\begin{algorithmic}[1]
\Procedure{Fitting Multiple Kernels}{}\newline
\textbf{Input:} A set of kernel matrices $\{K_d\}_{d=1}^D$, Data $(\by, \bX)$, $\lambda$ \newline
\textbf{Output:} Estimators $\bbeta$, $\{\balpha_d\}_{d=1}^D$\newline
$\#$ \texttt{Stage 1: Initialize parameters $\bbeta$ and $\{\balpha_d\}_{d=1}^D$}
\State{$\bbeta = (\bX^\top \bX)^{-1}\bX^\top \by$}
\For{$d = 1$ to $D$}
\State{$\balpha_d = (K_d + \lambda \bI)^{-1}(\by - \bX \bbeta)$}
\EndFor 
\newline
$\#$ \texttt{Stage 2: Iterative update}
\For{$step = 1$ to $max\_step$}
\State{$\bbeta = (\bX^\top \bX)^{-1}\bX^\top (\by-\sum_{d=1}^D K_d \balpha_d)$}
\For{$d = 1$ to $D$}
\State{$\balpha_d = (K_d + \lambda \bI)^{-1}(\by - \bX \bbeta-\sum_{d'\neq d} K_{d'} \balpha_{d'})$}
\EndFor
\EndFor 
\State{until convergence}
\EndProcedure
\end{algorithmic}
\end{algorithm}

\section{Projection matrices for multiple kernels}
Basically, for the problem \citep{buja_linear_1989},
\begin{align*}
    \by = \bX \bbeta + \sum_{d=1}^D K_d \balpha_d,
\end{align*}
define below projection matrices to the linear and the $d^{th}$ kernel as,
\begin{align*}
    H &= \bX (\bX^\top \bX)^{-1}\bX^\top, \\
    S_d &= K_d (K_d + \lambda \bI)^{-1}.
\end{align*}
Then we can compute the "projection matrices" to the overall kernel space $B$ as,
\begin{align*}
    A_d &= (\bI - S_d)^{-1}S_d,\\
    A &= \sum_{d=1}^D A_d,\\
    B &= (\bI + A)^{-1}A.
\end{align*}
Consequently, the final projections to the kernel effect space and the fixed effect space are,
\begin{align*}
    P_K &= (\bI -BH)^{-1}B(\bI -H),\\
    P_X &= H(\bI - P_K),
\end{align*}
such that $\bX \hat{\bbeta}=P_X \by$ and $\sum_{d=1}^D K_d \hat{\balpha}_d = P_K \by$.

\section{Derivation of the REML based test statistic}
\label{appendix_E}
\subsection{Derivation of the score test statistic}
In this section, we derive the score test statistic based on REML \citep{maity_powerful_2011}.\\ 
Denote $\bV(\btheta)=\sigma^2 \bV_\lambda=\sigma^2\bI+\tau \bK_\delta$, where $\btheta=(\delta, \tau, \sigma^2)$. The REML
\begin{align}
l_R(\mu, \lambda, \sigma^2|\by)=-\frac{1}{2}\Big[log\mid \sigma^2 \bV_\lambda \mid +(\by-\bmu)^T(\sigma^2 \bV_\lambda)^{-1}(\by-\bmu)+log \mid \sigma^{-2}\mathbf{1}^T\bV_\lambda^{-1}\mathbf{1} \mid \Big],
\end{align}
can be rewritten as
\begin{align}
l_R=-\frac{1}{2}\Big[log\mid \bV(\btheta) \mid +log \mid \mathbf{1}^T\bV(\btheta)^{-1}\mathbf{1} \mid +(\by-\bmu)^T\bV(\btheta)^{-1}(\by-\bmu)\Big]. \label{16}
\end{align}

Under $H_0: \delta=0$ (2.2.2), we set $\btheta_0=(0, \tau, \sigma^2)$ and
\begin{align*}
\bP_0(\btheta_0)=\bV(\btheta_0)^{-1}-\bV(\btheta_0)^{-1}\mathbf{1}[\mathbf{1}^T\bV(\btheta_0)^{-1}\mathbf{1}]^{-1}\mathbf{1}^T\bV(\btheta_0)^{-1}.
\end{align*}

Take the derivative of \eqref{16} with respect to $\delta$,
\begin{align}
\frac{\partial l_R}{\partial \delta}=&-\frac{1}{2}\Big[\frac{\partial log\mid \bV(\btheta) \mid}{\partial \delta}+\frac{\partial log \mid \mathbf{1}^T\bV(\btheta)^{-1}\mathbf{1} \mid}{\partial \delta}+ \frac{\partial (\by-\bmu)^T\bV(\btheta)^{-1}(\by-\bmu)}{\partial \delta}\Big]\nonumber \\
=&-\frac{1}{2}\Big[tr\big(\bV(\btheta)^{-1}\frac{\partial \bV(\btheta)}{\partial \delta}\big)+tr\big([\mathbf{1}^T\bV(\btheta)^{-1}\mathbf{1}]^{-1}\mathbf{1}^T\frac{\partial \bV(\btheta)^{-1}}{\partial \delta}\mathbf{1}\big)\nonumber \\
&+(\by-\bmu)^T\frac{\partial \bV(\btheta)^{-1}}{\partial \delta}(\by-\bmu) \Big]\nonumber \\
=&-\frac{1}{2}\Big[tr\big(\bV(\btheta)^{-1}\tau (\partial \bK_\delta)\big)-tr\big(\tau (\partial \bK_\delta)\bV(\btheta)^{-1}\mathbf{1}[\mathbf{1}^T\bV(\btheta)^{-1}\mathbf{1}]^{-1}\mathbf{1}^T\bV(\btheta)^{-1}\big)\nonumber \\
&-(\by-\bmu)^T\bV(\btheta)^{-1}\tau (\partial \bK_\delta)\bV(\btheta)^{-1}(\by-\bmu) \Big]\nonumber \\
=&\frac{1}{2}(\by-\bmu)^T\bV(\btheta)^{-1}\tau (\partial \bK_\delta)\bV(\btheta)^{-1}(\by-\bmu)\nonumber \\
&-\frac{1}{2}tr\Big[\tau (\partial \bK_\delta)\big[\bV(\btheta)^{-1}-\bV(\btheta)^{-1}\mathbf{1}[\mathbf{1}^T\bV(\btheta)^{-1}\mathbf{1} ]^{-1}\mathbf{1}^T\bV(\btheta)^{-1}\big]\Big], \label{17}
\end{align}
where $\partial \bK_\delta$ is the derivative kernel matrix whose $(i,j)^{th}$ entry is $\frac{\partial k_\delta(\bx, \bx')}{\partial \delta}$. If we further denote $\bK_0=\bK_\delta \mid_{\delta=0}$ and $\partial \bK_0=(\partial \bK_\delta)\mid_{\delta=0}$, we get the REML based score function of $\delta$ evaluated at $H_0$
\begin{align*}
S_{\delta=0}=\frac{1}{2}(\by-\bmu)^T\bV(\btheta_0)^{-1}\tau (\partial \bK_0)\bV(\btheta_0)^{-1}(\by-\bmu)-\frac{1}{2}tr[\tau (\partial \bK_0)\bP_0].
\end{align*}
To test for $H_0: \delta=0$, we propose to use the score-based test statistic
\begin{align}
\hat{T}_0=\hat{\tau}(\by-\hat{\bmu})^T\bV_0^{-1} (\partial \bK_0)\bV_0^{-1}(\by-\hat{\bmu}),
\end{align}
where $\bV_0=\hat{\sigma}^2\bI+\hat{\tau}\bK_0$.

\subsection{The null distribution of the test statistic}
For simplicity, we denote
\begin{align*}
\bV&=\bV(\btheta),\\
\bP=\bP(\btheta)=\bV^{-1}&-\bV^{-1}\mathbf{1}[\mathbf{1}^T\bV^{-1}\mathbf{1}]^{-1}\mathbf{1}^T\bV^{-1}.
\end{align*}

With similar derivation as \eqref{17}, for each $\theta_i \in \btheta=(\delta, \tau, \sigma^2)$, we have
\begin{align}
\frac{\partial l_R}{\partial \theta_i}=-\frac{1}{2}\Big[tr\big(\bP \frac{\partial \bV}{\partial \theta_i}\big)-(\by-\bmu)^T\bV^{-1}\big(\frac{\partial \bV}{\partial \theta_i}\big)\bV^{-1}(\by-\bmu)\Big]. \label{18}
\end{align}

From \citep{liu_semiparametric_2007} we know $\hat{\bmu}=[\mathbf{1}^T\bV^{-1}\mathbf{1}]^{-1}\mathbf{1}^T\bV^{-1}\by$. Plug it in \citep{lin_inference_1999}, and we obtain 
\begin{align*}
(\by-\bmu)^T\bV^{-1}=\by^T \big(\bI-\mathbf{1}[\mathbf{1}^T\bV^{-1}\mathbf{1}]^{-1}\mathbf{1}^T\bV^{-1}\big)^T\bV^{-1}=\by^T \bP.
\end{align*}

Then \eqref{18} becomes 
\begin{align*}
\frac{\partial l_R}{\partial \theta_i}=-\frac{1}{2}\Big[tr\big(\bP \frac{\partial \bV}{\partial \theta_i}\big)-\by^T \bP\big(\frac{\partial \bV}{\partial \theta_i}\big)\bP \by \Big].
\end{align*}

The second-order partial derivatives with respect to $\theta_i$ and $\theta_j$ are 
\begin{align}
\frac{\partial^2 l_R}{\partial \theta_i \partial \theta_j}=&-\frac{1}{2}\Big[tr\big(\frac{\partial \bP}{\partial \theta_j} \frac{\partial \bV}{\partial \theta_i}\big)+tr\big(\bP \frac{\partial^2 \bV}{\partial \theta_i \partial \theta_j}\big)+\by^T \bP\big(\frac{\partial \bV}{\partial \theta_i}\big)\bP \big(\frac{\partial \bV}{\partial \theta_j}\big)\bP \by \nonumber\\
&+\by^T \bP\big(\frac{\partial \bV}{\partial \theta_j}\big)\bP \big(\frac{\partial \bV}{\partial \theta_i}\big)\bP \by-\by^T \bP\frac{\partial^2 \bV}{\partial \theta_i \partial \theta_j}\bP \by \Big], \label{99}
\end{align}
where we have used the fact that
\begin{align*}
\frac{\partial \bP}{\partial \theta_j}=&-\bV^{-1}\frac{\partial \bV}{\partial \theta_j}\bV^{-1}+\bV^{-1}\frac{\partial \bV}{\partial \theta_j}\bV^{-1}\mathbf{1}[\mathbf{1}^T\bV^{-1}\mathbf{1}]^{-1}\mathbf{1}^T\bV^{-1}\\
&+\bV^{-1}\mathbf{1}[\mathbf{1}^T\bV^{-1}\mathbf{1}]^{-1}\mathbf{1}^T\bV^{-1}\frac{\partial \bV}{\partial \theta_j}\bV^{-1}\\&-\bV^{-1}\mathbf{1}\big([\mathbf{1}^T\bV^{-1}\mathbf{1}]^{-1}\mathbf{1}^T\bV^{-1}\frac{\partial \bV}{\partial \theta_j}\bV^{-1}\mathbf{1}[\mathbf{1}^T\bV^{-1}\mathbf{1}]^{-1}\big)\mathbf{1}^T\bV^{-1}\\
=&-\bP\frac{\partial \bV}{\partial \theta_j}\bP.
\end{align*}

Then \eqref{99} becomes 
\begin{align}
\frac{\partial^2 l_R}{\partial \theta_i \partial \theta_j}=&-\frac{1}{2}\Big[-tr\big(\bP \frac{\partial \bV}{\partial \theta_j}\bP \frac{\partial \bV}{\partial \theta_i}\big)+tr\big(\bP \frac{\partial^2 \bV}{\partial \theta_i \partial \theta_j}\big)+\by^T \bP\big(\frac{\partial \bV}{\partial \theta_i}\big)\bP \big(\frac{\partial \bV}{\partial \theta_j}\big)\bP \by \nonumber\\
&+\by^T \bP\big(\frac{\partial \bV}{\partial \theta_j}\big)\bP \big(\frac{\partial \bV}{\partial \theta_i}\big)\bP \by-\by^T \bP\frac{\partial^2 \bV}{\partial \theta_i \partial \theta_j}\bP \by \Big]. \label{19}
\end{align}

Since 
\begin{align*}
E(\bP \by \by^T)&=\bP [Var(\by)+(E\by)(E\by)^T]=\bP [\bV+\bmu \bmu^T]=\bP \bV, \\
\bP \bV \bP&=\bP[\bI-\mathbf{1}[\mathbf{1}^T\bV^{-1}\mathbf{1}]^{-1}\mathbf{1}^T\bV^{-1}]=\bP,
\end{align*}
we get
\begin{align*}
E\Big[\by^T \bP\big(\frac{\partial \bV}{\partial \theta_j}\big)\bP \big(\frac{\partial \bV}{\partial \theta_i}\big)\bP \by \Big]=&tr\Big(E\Big[\bP\big(\frac{\partial \bV}{\partial \theta_j}\big)\bP \big(\frac{\partial \bV}{\partial \theta_i}\big)\bP \by \by^T \Big] \Big)\\
=&tr\Big(\bP\big(\frac{\partial \bV}{\partial \theta_j}\big)\bP \big(\frac{\partial \bV}{\partial \theta_i}\big)\bP \bV \Big)\\
=&tr\Big(\bP\big(\frac{\partial \bV}{\partial \theta_j}\big)\bP \big(\frac{\partial \bV}{\partial \theta_i}\big)\Big),\\
E\Big[\by^T \bP\frac{\partial^2 \bV}{\partial \theta_i \partial \theta_j}\bP \by \Big]=&tr\Big(\bP\frac{\partial^2 \bV}{\partial \theta_i \partial \theta_j}\Big).
\end{align*}
Therefore, 
\begin{align*}
\bI_{\theta_i, \theta_j}=-E\Big[\frac{\partial^2 l_R}{\partial \theta_i \partial \theta_j}\Big]=\frac{1}{2}tr\Big(\bP\big(\frac{\partial \bV}{\partial \theta_j}\big)\bP \big(\frac{\partial \bV}{\partial \theta_i}\big)\Big).
\end{align*}

\section{Visualizing nonlinear interaction in Boston Housing Price}
\renewcommand\thefigure{F.\arabic{figure}}   
\setcounter{figure}{0} 
\label{appendix_F}

In this section we show an example of how to visualize the fitted interaction from a \code{cvek} model. Here we consider the \code{Boston} example in Section \ref{tutor:real}. We visualize the interaction effects by creating five datasets: Fix all confounding variables to their means, vary \texttt{lstat} in a reasonable range (i.e., from $12.5$ to $17.5$, since the original range of \texttt{lstat} in \texttt{Boston} dataset is $(1.73, 37.97)$), and respectively set \texttt{crim} value to its $5\%, 25\%, 50\%, 75\%$ and $95\%$ quantiles.

\begin{CodeChunk}
\begin{CodeInput}
> # first fit the alternative model
> formula_alt <- medv ~ zn + indus + chas + nox + rm + age + dis + 
+   rad + tax + ptratio + black + k(crim):k(lstat)
> fit_bos_alt <- cvek(formula = formula_alt, kern_func_list = kern_func_list, 
+                     data = Boston, lambda = exp(seq(-3, 5)))
> 
> # mean-center all confounding variables not involved in the interaction 
> # so that the predicted values are more easily interpreted
> pred_name <- c("zn", "indus", "chas", "nox", "rm", "age", 
+                "dis", "rad", "tax", "ptratio", "black")
> covar_mean <- apply(Boston, 2, mean)
> pred_cov <- covar_mean[pred_name]
> pred_cov_df <- t(as.data.frame(pred_cov))
> lstat_list <- seq(12.5, 17.5, length.out = 100)
> crim_quantiles <- quantile(Boston$crim, probs = c(.05, .25, .5, .75, .95))
> 
> # crim is set to its 5% quantile
> data_test1 <- data.frame(pred_cov_df, lstat = lstat_list, 
+                              crim = crim_quantiles[1])
row names were found from a short variable and have been discarded
> data_test1_pred <- predict(fit_bos_alt, data_test1)
> 
> # crim is set to its 25% quantile
> data_test2 <- data.frame(pred_cov_df, lstat = lstat_list, 
+                              crim = crim_quantiles[2])
row names were found from a short variable and have been discarded
> data_test2_pred <- predict(fit_bos_alt, data_test2)
> 
> # crim is set to its 50% quantile
> data_test3 <- data.frame(pred_cov_df, lstat = lstat_list, 
+                              crim = crim_quantiles[3])
row names were found from a short variable and have been discarded
> data_test3_pred <- predict(fit_bos_alt, data_test3)
> 
> # crim is set to its 75% quantile
> data_test4 <- data.frame(pred_cov_df, lstat = lstat_list, 
+                              crim = crim_quantiles[4])
row names were found from a short variable and have been discarded
> data_test4_pred <- predict(fit_bos_alt, data_test4)
> 
> # crim is set to its 95% quantile
> data_test5 <- data.frame(pred_cov_df, lstat = lstat_list, 
+                              crim = crim_quantiles[5])
row names were found from a short variable and have been discarded
> data_test5_pred <- predict(fit_bos_alt, data_test5)
> 
> # combine five sets of prediction data together
> medv <- rbind(data_test1_pred, data_test2_pred, data_test3_pred, 
+               data_test4_pred, data_test5_pred)
> data_pred <- data.frame(lstat = rep(lstat_list, 5), medv = medv, 
+                         crim = rep(c("5% quantile", "25% quantile", 
+                                      "50% quantile", "75% quantile", 
+                                      "95% quantile"), each = 100))
> data_pred$crim <- factor(data_pred$crim, 
+                          levels = c("5% quantile", "25% quantile", 
+                                     "50% quantile", "75% quantile", 
+                                     "95% quantile"))
> 
> data_label <- data_pred[which(data_pred$lstat == 17.5), ]
> data_label$value <- c("0.028%", "0.082%", "0.257%", "3.677%", "15.789%")
> data_label$value <- factor(data_label$value, levels = 
+                              c("0.028%", "0.082%", "0.257%", 
+                                "3.677%", "15.789%"))
>     
> ggplot(data = data_pred, aes(x = lstat, y = medv, color = crim)) + 
+     geom_point(size = 0.1) + 
+     geom_text_repel(aes(label = value), data = data_label, 
+                     color = "black", size = 3.6) + 
+     scale_colour_manual(values = c("firebrick1", "chocolate2", 
+                                    "darkolivegreen3", "skyblue2", 
+                                    "purple2")) + 
+     geom_line() + theme_set(theme_bw()) +
+     theme(panel.grid = element_blank(),
+           axis.title.x = element_text(size = 12), 
+           axis.title.y = element_text(size = 12), 
+           legend.title = element_text(size = 12, face = "bold"), 
+           legend.text = element_text(size = 12)) + 
+     labs(x = "percentage of lower status", 
+          y = "median value of owner-occupied homes ($1000)", 
+          col = "per capita crime rate")
\end{CodeInput}
\end{CodeChunk}

\begin{figure}[ht]
\begin{center}
\includegraphics[width=.8\columnwidth]{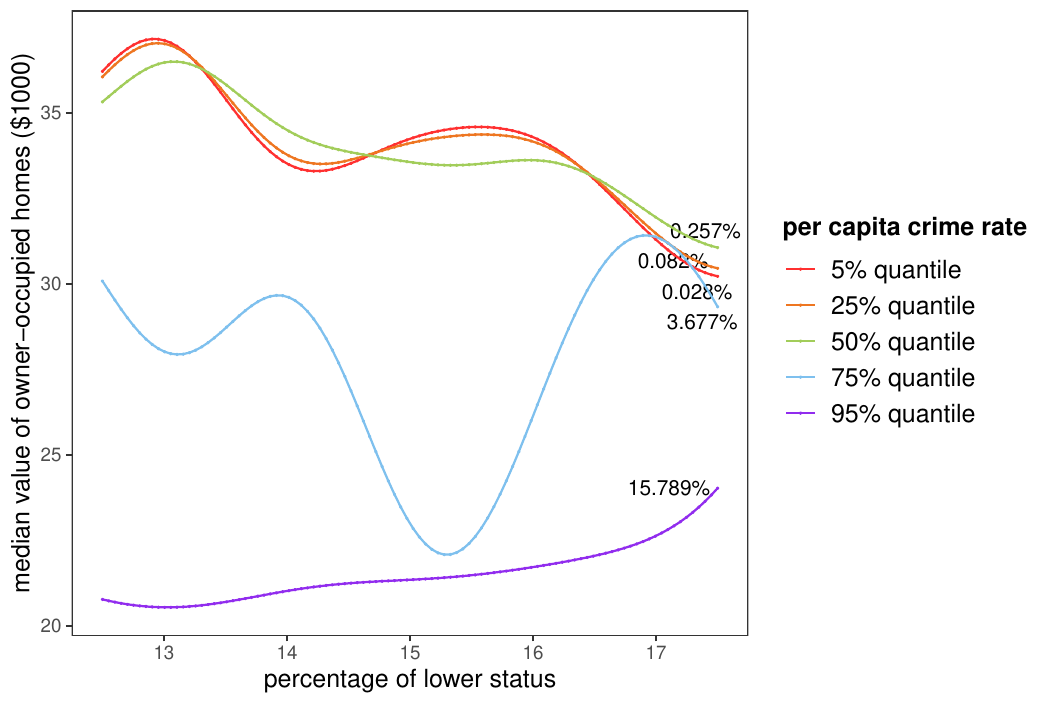} 
\caption{\texttt{medv} - \texttt{lstat} relationship under different levels of \texttt{crim}}
\label{fig:0}
\end{center}
\end{figure}

Figure \ref{fig:0} shows the \texttt{medv} - \texttt{lstat} relationship under different levels of \texttt{crim}. Numbers at the end of each curve indicate the actual values of \texttt{crim} rate (per capita crime rate by town) at the corresponding quantiles. From the figure we see that crime rate does impact the relationship between the local socioeconomic status v.s. housing price. Building on this code, user can continue to refine the visualization (e.g., by adding in confidence levels) and use it to improve the the model fit based on domain knowledge (e.g., by experimenting with different kernels / hyper-parameters).

\end{appendix}
\end{document}